\documentclass[usenatbib]{mn2e}

\usepackage{graphicx}
\usepackage[hypertex]{hyperref}

\usepackage{amssymb}

\def \eg {e.g.}
\def \ie {i.e.}
\def\spose#1{\hbox to 0pt{#1\hss}}
\def\ltsim{$\mathrel{\spose{\lower 3pt\hbox{$\sim$}}
        \raise 2.0pt\hbox{$<$}}$\thinspace}
\def\gtsim{$\mathrel{\spose{\lower 3pt\hbox{$\sim$}}
        \raise 2.0pt\hbox{$>$}}$\thinspace}
\newcommand{\thin }{\thinspace}

\newcommand{\lcdm}{$\Lambda$CDM}

\newcommand{\msun }{${\rm M_{\odot}}$}

\newcommand{\mfive}{${\rm M_{500}}$}
\newcommand{\cfive}{${\rm c_{500}}$}
\newcommand{\rfive}{${\rm R_{500}}$}

\newcommand{\model}{ScAM}

\newcommand{\rtwentyfive}{${\rm R_{2500}}$}

\newcommand{\src }{NGC\thin 720}
\newcommand{\srctwo}{RXJ\thin 1159+5531}
\newcommand{\zfe }{${\rm Z_{Fe}}$}

\newcommand{\suzaku}{{\em Suzaku}}
\newcommand{\chandra }{{\em Chandra}}

\newcommand{\xspec }{{\em Xspec}}

\newcommand{\xmm }{{\em XMM}}

\newcommand{\fnonthermal}{$f_{nth}$}

\newcommand{\fgas}{${\rm f_{g}}$}
\newcommand{\fg}{\fgas}

\newcommand\araa{{ARA\&A}}%
\newcommand\apj{{ApJ}}%
\newcommand\apjl{{ApJ}}%
\newcommand\apjs{{ApJS}}%
\newcommand\aap{{A\&A}}%
\newcommand\mnras{{MNRAS}}%
\newcommand\pasj{{PASJ}}%
\newcommand\nat{{Nature}}%
\newcommand\physrep{{Phys.~Rep.}}%
\defcitealias{humphrey06a}{H06}
\defcitealias{humphrey08a}{H08}
\defcitealias{humphrey11a}{H11}
\defcitealias{brighenti09a}{B09}

\title{Narrow Band X-ray Photometry as a Tool for Studying Galaxy and Cluster Mass Distributions}
\author[P.~J. Humphrey et al.]{\parbox{\textwidth}{Philip J. Humphrey,  David A. Buote} \vspace{0.4cm}\\
\parbox{\textwidth}{Department of Physics and Astronomy, University of California, Irvine, 4129 Frederick Reines Hall, Irvine, CA 92697-4575}}
\begin{document}
\maketitle
\begin{abstract}
We explore the utility of narrow band X-ray surface photometry as a tool for making 
fully Bayesian,
hydrostatic mass measurements of clusters of galaxies, groups and early-type galaxies.
We demonstrate that it is sufficient to measure the surface photometry with the \chandra\
X-ray observatory in only three (rest frame) bands (0.5--0.9~keV, 0.9--2.0~keV and 2.0--7.0~keV) in order
to constrain the temperature, density and abundance of the hot interstellar medium (ISM).
Adopting parametrized models for the mass distribution and radial entropy profile
and assuming spherical symmetry, we show that the constraints on the mass and thermodynamic 
properties of the ISM that are obtained by fitting data from all three bands simultaneously
are comparable to those obtained by fitting similar models to the temperature and density profiles derived
from spatially resolved spectroscopy, as is typically done. We demonstrate that 
the constraints can be significantly tightened when exploiting
a recently derived, empirical relationship between the gas fraction
and the entropy profile at large scales, eliminating arbitrary extrapolations at large radii.
This ``Scaled Adiabatic Model'' (\model) is well suited to modest signal-to-noise data, and 
we show that accurate, precise measurements of the global system properties are inferred
when employing it to fit data from even very shallow, snapshot X-ray observations.
The well-defined asymptotic behaviour of the model also makes it ideally suited for use in
Sunyaev-Zeldovich studies of galaxy clusters.
\end{abstract}
\begin{keywords}{Xrays: galaxies--- galaxies: elliptical and lenticular, cD--- galaxies: ISM--- dark matter--- methods: data analysis}
\end{keywords}
\section{Introduction}
The distribution of mass in galaxy clusters, groups and massive galaxies provides
a powerful tool for cosmological studies. Explicit predictions from our current
\lcdm\ cosmological paradigm for the number, size and radial mass 
distribution of 
dark matter halos can now be tested against high-quality constraints from 
studies employing lensing, Sunyaev-Zeldovich, stellar dynamics and, 
in particular, X-rays
\citep[\eg][]{voit05a,buote07a,mahdavi07a,gebhardt09a,vikhlinin09a,okabe10a,planck11a}. 
The relative distribution of dark and baryonic mass, coupled with the thermodynamic
state of the hot intracluster medium, similarly provides a unique
insight into the uncertain baryonic physics of galaxy formation, such as the role of feedback in 
shaping the nascent structures, and the complex interplay between adiabatic contraction 
and dynamical friction \citep[\eg][]{silk98a,gnedin04a,hopkins06a,abadi10a}. 

Spherical, hydrostatic X-ray techniques are an appealing method for measuring
such mass distributions due to their computational simplicity, given the isotropy
of the gas pressure tensor, and the small biases introduced by the spherical
approximation \citep[][and references therein]{buote11c}, particularly if the spherically
averaged mass profile is close to a singular isothermal sphere 
\citep{buote11b,churazov08a}. While the hot gas permeating the potential well 
is not expected to be exactly
hydrostatic, theoretical arguments and observational constraints suggest only
modest (\ltsim 30\%) biases on the inferred gravitating mass distribution, 
provided care is taken to study systems with relaxed X-ray morphologies 
\citep[\eg][]{buote95a,evrard96a,allen98a,nagai07a,piffaretti08a,churazov08a,mahdavi08a,fang09a,das10a,humphrey12c}. With the current generation of X-ray observatories, X-ray methods
are especially appealing as they can provide mass measurements over 
$\sim$3 orders of magnitude in virial mass, or more, and the radial mass 
distribution inferred can span a similarly large dynamical range in radius
\citep[\eg][]{buote07a,humphrey08a,humphrey11a,humphrey12a,wong11a}.

For spherically distributed hot gas in hydrostatic equilibrium, the 
radial mass profile can be uniquely inferred provided the gas density and 
temperature profiles are known \citep[\eg][]{mathews78a}. Prior to the 
launch of \chandra\ and \xmm, temperature profiles were typically sparsely
sampled, at best. In such circumstances, isothermality is a convenient 
approximation, since the gas temperature does not vary dramatically with radius.
This then implies a one-to-one relation between the gravitational potential and the 
density profile, and hence the surface brightness distribution, provided the abundance
profile is known (or, more usually, assumed to be flat).
For a \citet{king72a} gravitational potential, this leads to the 
ubiquitous ``isothermal $\beta$-model'' 
\citep[][]{cavaliere76a,cavaliere78a}.
The simple analytical form of the $\beta$-model has guaranteed its longevity as a convenient
ad hoc fitting function even though the underlying assumptions of the model are 
no longer believed to hold strictly \citep{arnaud09a}.

With the advent of \chandra\ and \xmm, spatially resolved spectroscopy has
largely superseded wide-band surface brightness photometry as a means for 
measuring the mass \citep[although see][]{fredericksen09a}, at least 
for high signal-to-noise (S/N) data \citep[\eg][]{voit05c,sun12a}. A
range of techniques have evolved for transforming the spectra into mass 
constraints \citep[][for a review]{buote11a}, 
most of which first entail fitting a single-phase plasma model to 
spectra from different regions of sky in order to obtain binned temperature (and,
possibly, density) profiles. This process often introduces correlations
between the binned temperature or density points, especially if 
deprojection techniques are employed or if coarser binning is used for
the temperature or abundance than the density. Care should be taken to
account for these, for example by using the full covariance
matrix to compute $\chi^2$ when model-fitting downstream,
rather than the common practice of just using the leading diagonal
\citep{pearson1900a,gould03a,humphrey11a}.
Even for gas that is strictly single phase in any
infinitesimal volume, temperature or abundance variations over the spectral 
extraction aperture violate the single phase approximation in that bin
and can lead to biases in the inferred temperature, abundance or
density profiles \citep[\eg][]{buote98c,buote00a,mazzotta04a,vikhlinin06c}.

Attempts to mitigate these issues have been made by modifying the spectral fitting
procedure. For example, \citet{eyles91a} and \citet{lloyddavies00a} fitted stacks of 
coarsely-binned, narrow-band images
(``data cubes'') by adopting parametrized models  for the temperature, 
abundance, and either the gas density or gravitating mass profiles.
(In the latter case, the gas density profile was then 
derived under the hydrostatic approximation.)
Given the physical state of the gas as a function of position predicted by
this model, spectra were generated in a series of shells that were, 
in turn, projected along the line of sight
and fitted directly to the data cube. This circumvents the intermediate step
of measuring the binned temperature profile.
 Similar approaches, albeit emphasizing the simultaneous fitting of 
full-resolution spectra obtained from concentric annuli,
were advocated by  \citet{pizzolato03a} and \citet{mahdavi07a}.

In objects with lower surface brightness  it is often impossible to
obtain sufficient photons to enable high-quality spectral analysis 
in as many bins as required. In these cases, it is common practice to 
measure coarse, global quantities such as the emission-weighted luminosity 
(determined, for example, from a $\beta$-model fit), temperature
or $Y_X$ \citep[the product of temperature and gas mass:][]{kravtsov06a}, and apply
scaling relations to transform these into mass estimates
\citep[\eg][]{voit05c,sun12a}. The calibration of these scaling relations is generally empirical,
and, to be reliable, requires high resolution spectroscopy of objects similar 
to those under scrutiny. Any given object cannot, in practice, be
guaranteed to obey these relations, and assuming this behaviour can, therefore,
 restrict discovery space. 

As a compromise between these two extremes (global scaling relations and 
spatially resolved spectroscopy), we propose a powerful alternative
method, {\em narrow band photometry}, \ie\ the simultaneous fitting 
of radial X-ray surface brightness profiles in multiple Pulse Height 
(PHA) bands, as an appropriate choice for modest-S/N data.
This involves first adopting models for the three-dimensional 
gas temperature, density and abundance profiles, which are then used to derive 
the predicted surface  brightness distribution in each band.
While this has philosophical similarities to the 
data cube fitting of \citet{eyles91a}, in this paper, we demonstrate that
only a limited number of energy bands are necessary to constrain the 
mass of the system. By emphasizing radial surface brightness photometry, 
the treatment of the 
background is also simplified. If the source and background models 
are constrained together, spectral based methods generally require the 
arbitrary parametrization of the background spectrum, as well as usually 
employing assumptions over its spatial variation 
\citep[\eg][]{buote04c,humphrey11a,liu12a}.
In our approach, we demonstrate that minimal assumptions 
need to be made about the spectral shape of the background.

An essential ingredient of our approach is a physical model for the 
three dimensional distribution of the gas density and temperature. 
In our recent work, we have demonstrated the utility of an 
entropy-based model, in which the gas density and temperature profiles
are inferred self-consistently in the hydrostatic approximation
from a model for the (non-gas) gravitating mass and the gas entropy 
\citep{humphrey08a,humphrey09d}. Part of the advantage of this parametrization
is that it allows convective stability (a necessary condition for hydrostatic
equilibrium) to be rigorously enforced, which imposes additional important
constraints on the mass distribution \citep[\eg][]{fabian86a}. 
Similar, entropy-based approaches have been independently explored in the 
context of Sunyaev-Zeldovich studies \citep{allison11a}, and in a more
non-parametric manner by \citet{cavaliere09a}.

We have found that a powerlaw model (with
arbitrary, multiple breaks) and a central plateau is sufficient to fit 
typical entropy profiles measured from high-quality, spatially resolved spectroscopy
\citep[\eg][]{cavagnolo09a,humphrey12a}. When using such a model, a major
source of uncertainty is the behaviour of the entropy at the largest physical scales,
where the S/N of the data is lowest, which can translate into errors on the recovered
global properties, such as the gas fraction \citep[\eg][]{humphrey11a}. 
In addition to the usual powerlaw parametrization, in this paper, we therefore
also explore a class of model that mitigates this uncertainty by exploiting,
in a somewhat revised form, the recently
discovered correlation between the gas fraction and entropy profile, which spans 
a wide range of halo mass
\citep{pratt10a,humphrey11a,humphrey12a,humphrey12b}.

We discuss the practical implementation of narrow band photometry in X-ray analysis in
\S~\ref{sect_narrow_band}. In \S~\ref{sect_entropy_model} we illustrate how this 
technique can be used in conjunction with the entropy-based, hydrostatic gas
model to constrain the gravitating mass over a range of mass scales by employing
simulations tailored to match two well-studied objects from the literature. 
Finally, in \S~\ref{sect_entropy_scaling}, we show how additional constraints on the entropy profile
lead to a new class of model that is optimal for fitting low S/N data, and 
reach our conclusions in \S~\ref{sect_discussion}.

\section{Narrow band X-ray photometry} \label{sect_narrow_band}
In this section, we discuss the practical implementation of narrow band photometry
as a tool for X-ray astronomy. 
\subsection{Implementation}
Let us define the X-ray surface brightness (in $counts\ s^{-1}\ arcmin^{-2}$) 
measured for a given object in a particular PHA (energy) channel band 
($i_0$--$i_1$) over some region of the detector $k$
\begin{eqnarray}
SB^k_{i_0,i_1} & = & \frac{\sum_{i=i_0}^{i_1} C_i^k}{A_k} \label{eqn_sb_definition}
\end{eqnarray}
where $A_k$ is the area of the region measured in square arc minutes.  $C_i^k$ is the 
photon count rate detected in PHA channel $i$ and region $k$. Assuming the source emission
comes from an optically thin plasma, we can write (see Appendix~\ref{sect_SB}):
\begin{eqnarray}
C_i^k & = & \int \int dx dy\delta_k(x,y) \int_{2\pi} d\Omega \int dZ \int d\nu^\prime \bigg[ \nonumber \\
      & &  \hspace{60pt} \frac{\epsilon_{\nu^\prime}({\bf \Omega},Z)}{(1+z)^3} T_i(\nu,x,y,{\bf \Omega}) \bigg] \label{eqn_ci_definition}
\end{eqnarray}
where $x$ and $y$ are 
detector coordinates (in arc minutes), $\delta_k(x,y)=1$ if $(x,y)$ is within region $k$
or is 0 otherwise, $\Omega$ is the solid angle, $Z$ is the line-of-sight coordinate to the 
object, $\nu^\prime$ is the photon frequency in the source rest frame, $\epsilon_{\nu^\prime}$
is the (rest frame) count rate of photons emitted by the source per unit volume, per unit 
frequency range, $z$ is the source redshift, and $T_i(\nu,x,y,\Omega) d\Omega dx\ dy$ is the
likelihood that a photon of energy $\nu=\nu^{\prime}/(1+z)$ incident at angle $\Omega$
will be detected in PHA channel $i$ at detector coordinate $(x,y)$. $T_i$ contains information
about the telescope and detector sensitivity, point spread function and spectral redistribution
function.

Our aim is to invert  Eqn~\ref{eqn_ci_definition} to infer 
$\epsilon_{\nu^\prime}$ as a function of position in the object, and thus the physical state of the
emitting medium. Forward fitting is a powerful way to achieve this. This entails
adopting
a reasonable parametrization for $\epsilon_{\nu^\prime}$ as a function of ($\Omega,Z$), and 
evaluating Eqn~\ref{eqn_ci_definition} numerically. The adopted parameters are adjusted until
good agreement is found with the measured $SB^k_{i_0,i_1}$. As we discuss below (\S~\ref{sect_hardness}),
it is generally insufficient to use a single photometric band $i_0$--$i_1$, and so multiple
surface brightness profiles ($SB_k$) will need to be fitted simultaneously. 
To simplify the problem,
spherical symmetry is generally assumed. In Appendix~\ref{sect_SB} we outline a practical 
scheme for evaluating the integral efficiently in that case. 
\begin{figure*}
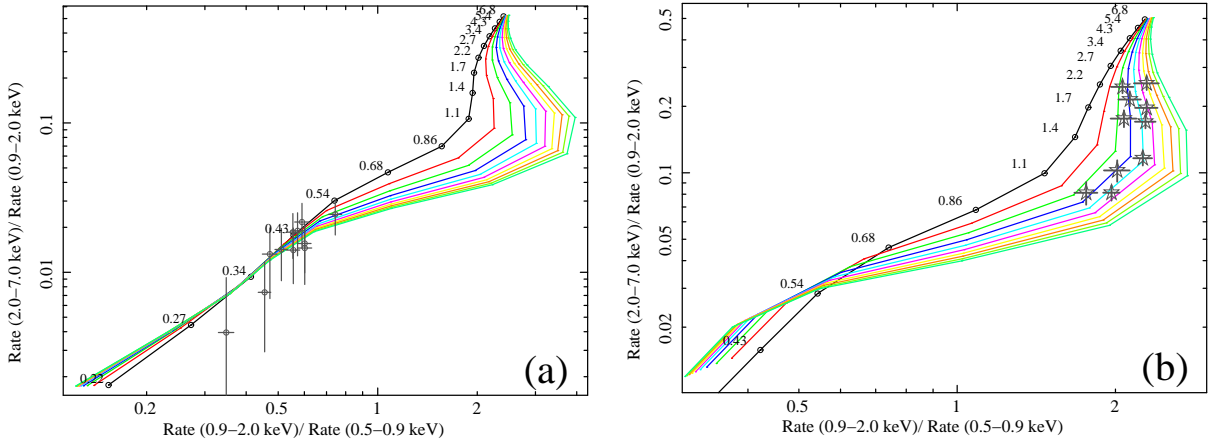

\includegraphics[height=3.2in,angle=270]{hardness_n720.ps}
\includegraphics[height=3.2in,angle=270]{hardness_rxj1159.ps}
\caption{Hardness-hardness plot for an APEC thermal plasma, for two different representative 
\chandra\ ACIS-S3 observations. Each line corresponds to the locus of constant abundance
(varying from 0.1 times Solar: black, to Solar: blue-green, in 0.1 increments) as kT varies
(marked at various intervals). 
We assumed Solar abundance ratios \citep{asplund04a} and adopted redshifts of 0 (a) and
0.081 (b); all energy bands are in the observer's frame.  For illustration purposes, we show
the measured hardness ratios in a series of different radial bins for two simulated observations,
a and b, which are tailored to resemble the real systems of \src\ and \srctwo\ 
\citep{humphrey11a,humphrey12a}. For both systems, the hardness ratios are sufficiently defined
to constrain the temperature and, for object b, the abundance can be determined as well.} \label{fig_hardness}
\end{figure*}

\subsection{Required spectral resolution} \label{sect_hardness}
Directly fitting a coronal plasma model to the full resolution spectrum of the hot 
gas provides the most reliable means for measuring the temperature and
metallicity. Provided the abundance ratios of key species with respect to Fe are known
(for example, if they match the Solar abundance pattern), however, hardness 
ratios (\ie\ the ratio of the detected count-rates in two energy-bands) can also be used
to extract temperature and abundance information. 

For representative \chandra\ ACIS-S3 observations, we show in Fig~\ref{fig_hardness}
the loci in hardness-hardness space of APEC thermal plasma models having 
different temperatures and abundances. We have carefully chosen the energy-bands under
scrutiny to maximize the separation of the models, but it is clear that two hardness ratios 
(\ie\ three energy bands) are sufficient to constrain the gas temperature and abundance 
(and hence density) in these observations. Still, this approach requires careful calibration 
of the hardness ratios on an observation by observation basis, taking into account
the spatial variation of the instrumental response, the Galactic absorbing column and the 
redshift of the source.
For reference, we also show the corresponding hardness ratios in a series of radial
bins for two simulated datasets (discussed in \S~\ref{sect_simple_entropy_performance}),
illustrating that, for real data, this provides a feasible means to measure the gas
properties in multiple annuli.

\subsection{Isothermal $\beta$-model}
For an optically thin coronal plasma, we expect $\epsilon_{\nu^\prime}$ to depend only
on the photon frequency ($\nu^\prime$), the gas density ($\rho_g$), temperature (expressed
in energy units as kT), 
and abundances ($Z_l$) of key species (denoted here as $l$).
As a simple illustration of the method, we therefore consider the case of such a 
plasma. Assuming spherical symmetry and isothermal gas with a $\beta$-model density
profile,
\begin{equation}
\epsilon_{\nu^\prime} = n_{e,0}n_{H,0}  \left( 1+\frac{r^2}{r_c^2} \right)^{-3\beta} \sum_l 4\pi \Lambda_l(kT) Z_l
\label{eqn_beta}
\end{equation}
where $n_{e,0}$ and $n_{H,0}$ are the central electron and Hydrogen number densities, respectively,
$\Lambda_l$ relates the gas emissivity at frequency $\nu^\prime$ 
to the temperature for the species $l$, which has abundance 
$Z_l$. Here $r_c$ and $\beta$ are the 
standard parameters of the $\beta$-model  (\ie\ the ``core radius'' and $\beta$ parameter).
In Appendix~\ref{sect_SB}, we 
derive a generalized expression for the surface brightness in terms of $\epsilon_{\nu^\prime}$.
Substituting Eqn~\ref{eqn_beta} into Eqn~\ref{eqn_n_nu_j1} and integrating, we obtain:
\begin{eqnarray} 
SB_{i_0,i_1}^{k} \simeq  F \sum_j N_j \sum_{i=i_0}^{i_1} \sum_l  Z_l \int d\nu^\prime\overline{R}^k_i(\nu) \overline{P}_{\nu}^{j\rightarrow k} A_{eff}^{j\rightarrow k} \Lambda_l (kT)\label{eqn_beta_model}
\end{eqnarray}
where 
\begin{eqnarray}
F &  = &  \frac{n_{e,0}n_{H,0} }{(1+z)^3} \left( \frac{\Gamma(0.5) \Gamma(3\beta-0.5)}{\Gamma(3\beta)} \right) \frac{r_c^3}{D_A^2} \\
 N_j &  = &  \left[\left(1+\frac{R_{j+1}^2}{r_c^2} \right)^{3\beta-\frac{3}{2}} - \left(1+ \frac{R_{j}^2}{r_c^2} \right)^{3\beta-\frac{3}{2}} \right] 
\end{eqnarray}
if $\beta \ne 0.5$. Here, $\Gamma$ is the gamma function. If $\beta=0.5$, then 
\begin{eqnarray}
 N_j  & =  & \ln \left(\frac{r_c^2+R_{j+1}^2}{r_c^2+R_j^2} \right) 
\end{eqnarray}
The integral in Eqn~\ref{eqn_beta_model} does not depend on $\beta$ or $r_c$, and instead only
depends on the (known) PHA bands used for surface photometry, the (known) annuli definitions
and the (unknown) gas temperature. By fitting surface brightness profiles
in multiple bands simultaneously, therefore, it is possible to constrain the temperature and abundance,
along with $\beta$ and $r_c$. For speed it is often convenient to adopt a linear approximation
for $\Lambda_l$; \ie:
\begin{eqnarray}
\Lambda_l (kT) & \simeq & \Lambda_l (kT_m) \frac{T_{m+1}-T}{T_{m+1}-T_m} + \Lambda_l (kT_{m+1}) \frac{T-T_m}{T_{m+1}-T_m} \nonumber
\end{eqnarray}
where $m$ is chosen such that $kT_m \le kT < kT_{m+1}$. 
(We note that this same approximation is used within the \xspec\ X-ray 
spectral fitting package for the computation of the APEC plasma emissivity.) This approximation 
means that 
kT can be taken outside of the integral in Eqn~\ref{eqn_beta_model}, allowing the
integrals over $\nu^\prime$ to be pre-computed for speed.

\subsubsection{Performance of the model} \label{sect_fit_beta}
To illustrate the utility of this approach, we fitted the model to realistic,
simulated \chandra\ surface brightness profiles. We assumed the hot gas
was distributed as a $\beta$-model with $\beta=0.389$, $r_c=0.42$~kpc (which 
corresponds to 3.4\arcsec, adopting a distance of 25.7~Mpc) and a central
gas density of $2.08\times 10^{-25} g\ cm^{-3}$. These parameters were chosen to
be similar to the properties of the isolated, Milky Way-mass elliptical galaxy 
NGC\thin 720 \citep{humphrey11a}. Adopting 1~keV for the 
gas temperature and abundances which are 0.5 times Solar \citep{asplund04a},
we simulated an artificial \chandra\ events file, using the algorithm
outlined in \citet{humphrey12c}. Briefly, we computed the gas emissivity
in a series of (R,z) bins in cylindrical polar coordinates, assuming an
APEC thermal plasma model. Folding in the effective area and spectral 
response function of \chandra, we then simulated a corresponding spectrum
from each region for a 100~ks exposure time, and the resulting 
photons were projected onto the sky
to build up the event list. In \citet{humphrey12c}, we demonstrated the 
reliability of this method for producing artificial data.
The resulting events file contains $\sim$20000
photons from the hot gas within a 4\arcmin\ region. We added a background component with a 
constant surface brightness profile and a plausible spectrum (two soft
APEC thermal plasma components, and a powerlaw, normalized to match
the sky background in the vicinity of NGC\thin 720). For simplicity,
we ignored the spatial dependence of the effective area and spectral response,
and ignored the spatial point spread of the telescope.

We extracted surface brightness profiles in three energy bands,
0.5--0.9~keV, 0.9--2.0~keV and 2.0--7.0~keV, corresponding to those 
shown in Fig~\ref{fig_hardness}. We fitted a model comprising the isothermal
$\beta$-model described above (Eqn~\ref{eqn_beta_model}) for the hot
gas and a constant background component in each band. We allowed 
$\beta$, $r_c$, the central gas density ($\rho_g$), kT and the gas abundance to be fitted freely.
Fits were performed by 
minimizing the C-statistic, which is appropriate given the Poisson distributed nature
of the data \citep[\eg][]{humphrey09b,cash79a} using dedicated software based around the 
MINUIT software library\footnote{\href{http://lcgapp.cern.ch/project/cls/work-packages/mathlibs/minuit/index.html}{http://lcgapp.cern.ch/project/cls/work-packages/mathlibs/minuit/index.html}}.
By visual inspection, the overall fit was good and each parameter was recovered accurately,
specifically within 2-$\sigma$ of the true value (kT=$0.99\pm0.01$~keV, 
\zfe=$0.43\pm0.05$,
$\beta=0.387\pm0.003$, $r_c=0.40\pm0.03$~kpc, and 
$\rho_g=(2.2\pm0.2)\times 10^{-25}g\ cm^{-3}$). The ability to constrain
both gas temperature and abundance by using these three energy bands is 
unsurprising given the temperature of the gas (Fig~\ref{fig_hardness}).

\section{Mass analysis} \label{sect_entropy_model}
\begin{figure*}
\includegraphics[width=7in]{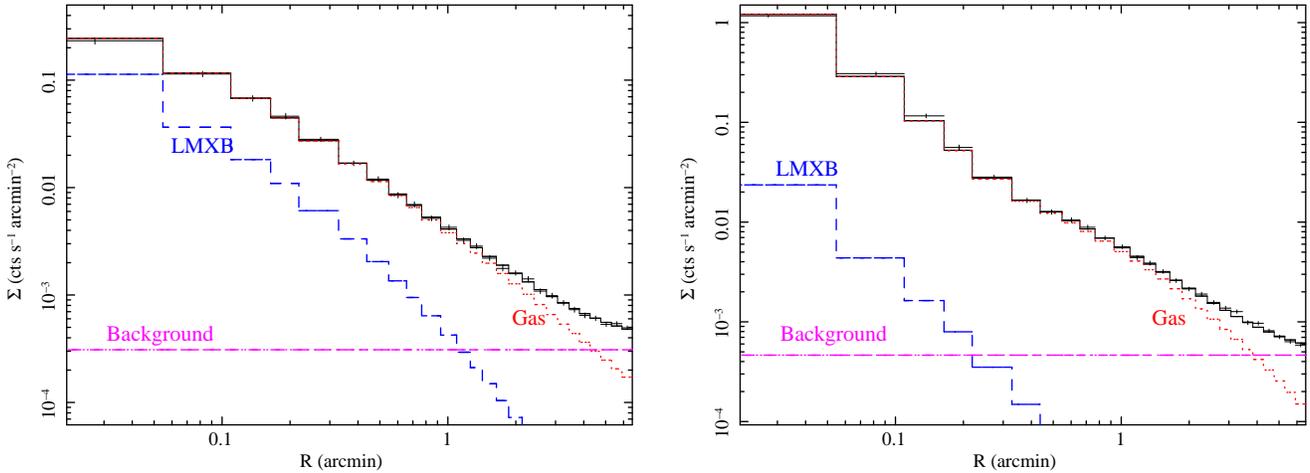}
\caption{0.9--2.0~keV surface brightness profiles for the simulated datasets a (left panel) and b
(right panel). Overlaid (black solid lines) are the best-fitting entropy-based hydrostatic model,
and the contribution to this model from hot gas (red dotted line), unresolved X-ray binaries 
(blue dashed line) and the background (magenta dash-dotted line). The overall fits are good.} \label{fig_rxj1159_sb} \label{fig_n720_sb}
\end{figure*}
We next consider a more physically motivated model for the surface brightness profile,
based on a self-consistent solution to the equation of hydrostatic equilibrium for 
the density and temperature profiles.
\subsection{Method}
If the gas is spherically
distributed and in an equilibrium state  (although not necessarily static), 
we can write \citep{humphrey08a,humphrey12c,fang09a,buote11a}:
\begin{eqnarray}
\overline{r}^2 \overline{S}^\frac{3}{5} \frac{d\overline{\zeta}}{d\overline{r}} = 
-\frac{2}{5} \left( \frac{G M_{500}}{r_{500}}\right) \left( \frac{\mu m_H}{kT_{ref}}\right) \left( \overline{M_g} + \overline{M_{ng}} -\overline{M_{eff}}\right) \label{eqn_zeta}\\
\frac{\overline{S}^\frac{3}{5}}{\overline{r}^2} \frac{d\overline{M_g}}{d\overline{r}} = 
\left( \frac{4\pi r_{500}^3}{M_{500}} \right) \left( \frac{k T_{ref}}{\mu m_H S_{500}}\right)^\frac{3}{2} \overline{\zeta}^\frac{3}{2}  \label{eqn_mg}
\end{eqnarray}
where $G$ is the Universal gravitational constant, $r_{500}$ is the radius within which the 
mean mass density of the system is 500 times the critical density of the Universe, $M_{500}$
is the mass within $r_{500}$, $\mu$ is the mean molecular weight ($\sim 0.62$ for a fully ionized
plasma), $m_H$ is the mass of hydrogen, $kT_{ref}$ is an arbitrary reference energy (typically
1~keV), $\overline{r}$ is the radius from the centre of the object divided by $r_{500}$,
$\overline{M}_g$ is the gas mass (divided by $M_{500}$) enclosed within radius $\overline{r}$,
$\overline{M}_{ng}$ is the enclosed non-gas mass, $\overline{M}_{eff}$ is the 
``effective mass'' codifying the bulk gas motions, including
rotation (in the notation of \citealt{humphrey12c}, $\overline{M}_{eff}=r v_{eff}^2/(G M_{500})$. 
For pure rotation, $v_{eff}$ is the spherically averaged rotation velocity at radius r).
We define $\overline{S}=\rho_g^{-\frac{2}{3}} kT/(\mu m_H S_{500} (1-f_{nth}))$, corresponding
to the usual astronomers' entropy proxy,
where $f_{nth}$ is the nonthermal pressure fraction (including the effects of magnetic 
pressure, cosmic ray pressure and turbulence), and 
\begin{eqnarray}
S_{500}& =& 150 \left( \frac{M_{500}}{10^{14}M_\odot f_{b,U} E_z h_{70}^2}\right)^\frac{2}{3} \left( \frac{2+\mu}{5\mu} \right)^\frac{2}{3} \mu^{-1} m_H^{-\frac{5}{3}} \nonumber \\
E_z& =& \Omega_m (1+z)^3+(1-\Omega_m)
\end{eqnarray}
where $f_{b,U}$ is the Cosmological baryon fraction (0.17; \citealt{dunkley09a}), $h_{70}$ is the
Hubble constant in units of $70 km\ s^{-1} Mpc^{-1}$, and $\Omega_m$ is the cosmological matter 
density parameter. We also define $\overline{\zeta}=\left( \rho_g kT/(\mu m_H (1-f_{nth})
\right)^\frac{2}{5} \mu m_H S_{500}^\frac{3}{5}/kT_{ref}$, which is proportional to the thermal
gas pressure to the power of $2/5$.
The true values of $r_{500}$ and $M_{500}$ 
(which include the gas mass) are, technically, not known prior to solving the equations for
$M_g$. However, since the scaling factors are arbitrary, we instead adopted the 
values of $r_{500}$ and $M_{500}$ appropriate for the dark matter halo alone, which can be 
computed analytically for a \citet[][hereafter NFW]{navarro97} profile.

By adopting parametrized models for $\overline{S}$, $\overline{M_{ng}}$, \fnonthermal, and 
$\overline{M_{eff}}$, we can solve Eqns~\ref{eqn_zeta}--\ref{eqn_mg} numerically
(\eg\ with a fourth order Runge-Kutta method) for any given set of input parameters,
and hence determine the radial density and temperature profiles. 
A reasonable model for $\overline{M}_{ng}$  comprises a 
stellar mass component (assuming mass follows light, this can be derived by deprojecting the 
stellar light; \eg\ \citealt{binney90a,cappellari02a,humphrey09d,humphrey12b}) 
with adjustable M/L ratio, 
a central supermassive black hole, and a dark matter halo (\eg\ NFW).  If the gas is 
close to hydrostatic, $f_{nth}\simeq 0$ and $M_{eff}\simeq 0$, leaving it necessary only 
to adopt a parametrized form for $\overline{S}$, and provide two boundary conditions
(on $\overline{M}_g$  and $\overline{\zeta}$).

We begin the integration from a small radius, $\overline{r}_0$, which is sufficiently
small that we can impose the boundary condition $\overline{M}_g(<\overline{r}_0)=0$,
\ie\ we assume there is no gas within $\overline{r}_0$. The boundary condition on 
$\overline{\zeta}$ is an adjustable parameter, which significantly affects the shape of the
resulting gas temperature and density distribution. In particular, we found that 
the temperature profile shape can change from radially falling to radially rising 
(a ``cool core'' profile) by increasing $\overline{\zeta}(\overline{r}_0)$. That changing the 
pressure boundary condition can produce such diverse behaviour is unsurprising 
\citep{mathews03a}. The outer radius of integration is arbitrary, but it should 
be sufficiently large that it accounts for almost all of the emission that will
be projected into the line-of-sight (see \S~\ref{sect_SB}). We advocate some fraction
(or multiple) of the virial radius as the optimal choice here, although it is important
to understand that this requires that  the gas remains approximately spherically 
distributed and hydrostatic (or, at least, accurately described by the adopted forms for 
\fnonthermal\ and $\overline{M}_{eff}$)  out to this scale.

\subsection{Simple entropy profile} \label{sect_simple_entropy}
In our recent work, we have assumed that the entropy depends on radius as a 
(multiply broken) powerlaw, plus a constant 
(see \citealt{humphrey08a,humphrey12a,humphrey12c}), \ie\
 \begin{eqnarray}
\overline{S} & = s_0 + s_1 \overline{r}^{\alpha_1} & (\overline{r}<\overline{r}_{b1}) \nonumber \\
  & = s_0 + s_1 \overline{r}_{b1}^{\alpha_1-\alpha_2} \overline{r}^{\alpha_2} & (\overline{r}_{b1} \le \overline{r} < \ldots) \label{eqn_powerlaw}
\end{eqnarray}
and so on, adding in as many breaks as required. The parameters $s_0$, $s_1$, $\alpha_1$,
$\alpha_2$ and $r_{b1}$ are adjustable. 
Empirically, observed entropy profiles are often well parametrized by 
similar models 
\citep[\eg][]{mahdavi05a,donahue06a,jetha07a,finoguenov07a,gastaldello07b,sun09a,cavagnolo09a,johnson09b,pratt10a,humphrey12a,werner12a}. 
Theoretically, powerlaw-like entropy profiles are expected from adiabatic 
structure formation models \citep{tozzi01a,voit05b}, which may be modified by
feedback, resulting in profiles that resembles (multiply) broken powerlaw
distributions, with a constant offset \citep[\eg][]{voit05a,mccarthy10a}.

\subsection{Metallicity profile} \label{sect_abundance}
In order to convert the measured gas density and temperature into a surface brightness profile,
it is necessary to know the three dimensional radial dependence of the abundance pattern. In practice,
one can adopt a simple parametrization, the free parameters of which can be fitted along with the 
parameters controlling the mass model and thermodynamic properties of the gas. 
Unfortunately there is no generally accepted functional form for such a distribution, and so
we recommend experimentation with different metallicity profiles as a means of 
exploring the sensitivity of the results to the details of this assumption. For the purpose of validating the 
surface brightness model as an adequate tool for fitting, it is convenient to adopt flat (\ie\ constant)
metallicity profiles in this paper. We allowed only the total (Fe) abundance to be fitted, while fixing the 
abundance ratios of other metals with respect to Fe to be at their Solar
values \citep{asplund04a}. We return to the problem of a radially varying abundance profile in 
\S~\ref{sect_discuss_abundance}.


\subsection{Performance of the model} \label{sect_simple_entropy_performance}
\begin{figure*}
\includegraphics[height=7in,angle=270]{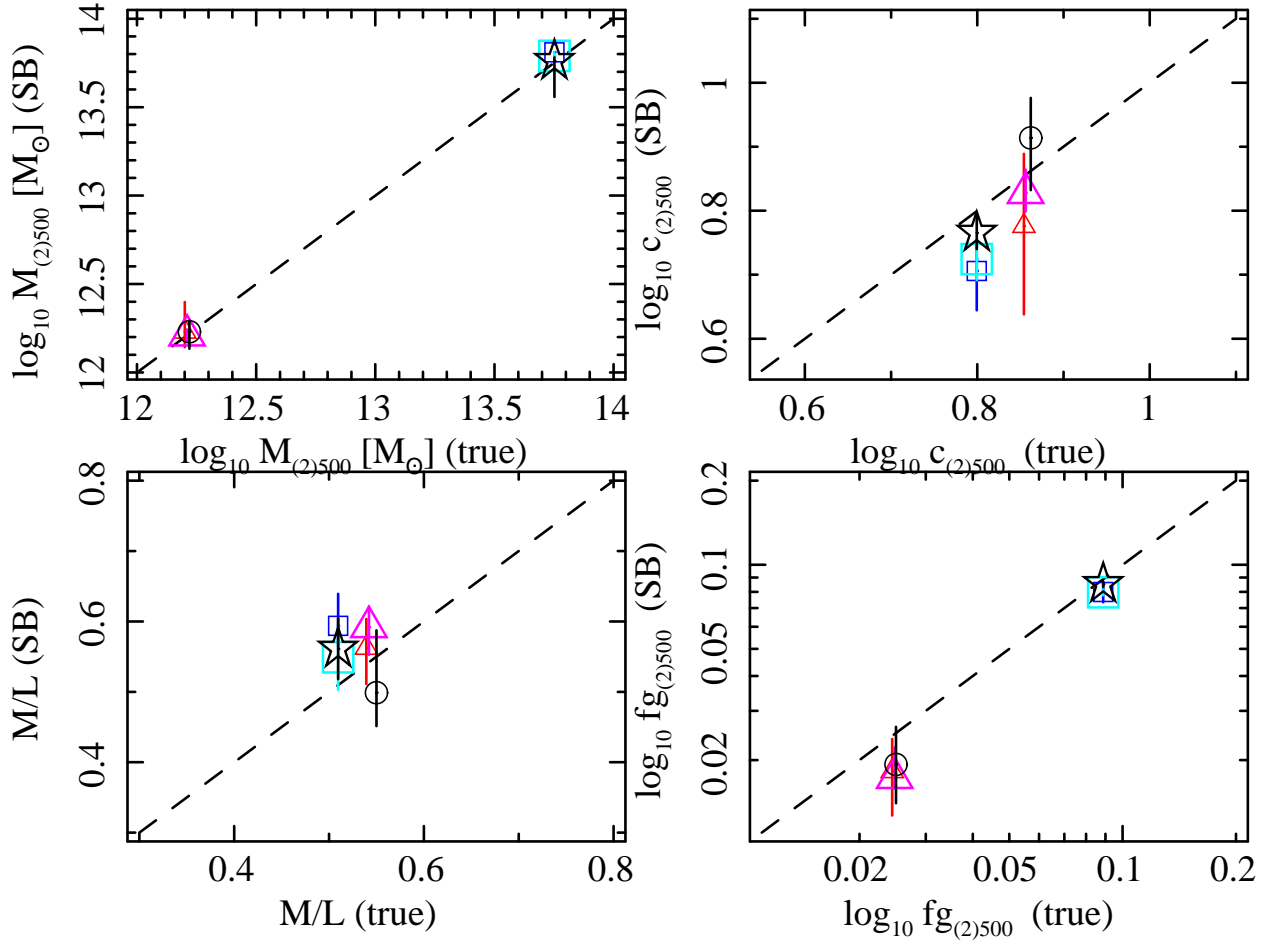}
\caption{Accuracy of various physical parameters obtained from narrow-band photometric fits.
We compare total mass (M), halo concentration (c), stellar mass-to-light ratio (M/L) and
gas fraction ($f_g$) measurements.
For simulated 100~ks datasets a (triangles) and b (squares), the parameters derived from fitting the 
hydrostatic model with a ``simple entropy profile'' (small symbols) and the \model\ model (large 
symbols) are shown on the y-axes, and the true value is shown as the x-axis. The dashed lines
indicate equality ($y=x$). Comparisons are made between parameters at either \rtwentyfive\ (for 
\src) or \rfive\ (for \srctwo), indicated by the 2(500) notation.
We find that the model fit results are both precise and accurate, being
within $\sim 2 \sigma$ of the true value in each case. As expected, the constraints obtained with 
the \model\ model are tighter. In addition, we also show the results of fitting the \model\ model
to a 5~ks snapshot observation of object a (large circles), demonstrating the utility of this 
model with shallow exposures. For clarity, these points are shown slightly displaced in the 
x-direction.
We show (large stars) the results of fitting the \model\ model
to object b, but allowing the Fe abundance profile to peak in the centre (see \S~\ref{sect_discuss_abundance}).} 
\label{fig_compare_derived}
\end{figure*}
\begin{figure*}
\includegraphics[width=7in]{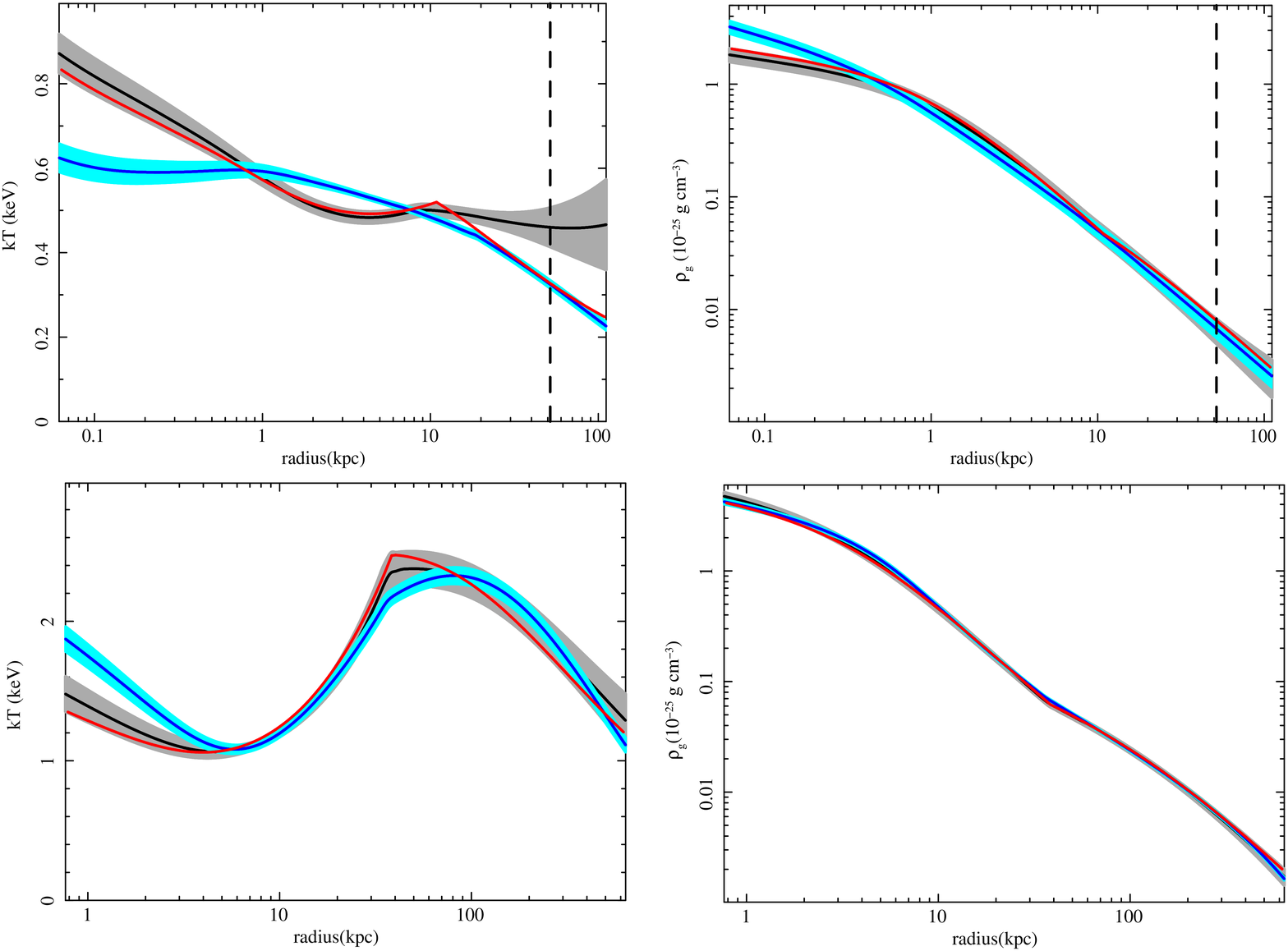}
\caption{Three dimensional temperature and density profiles for simulated objects a (top panels) and b
(lower panels),
obtained from fitting the hydrostatic narrow band photometric model with a simple entropy 
profile (grey region) and the \model\ model (blue region) to the 100~ks simulated datasets. The red line shows the true profile.
The shaded regions indicate 1-$\sigma$ confidence regions.
The dashed vertical line in the upper panels corresponds to $\sim$7\arcmin, demarcating the 
region for which surface brightness fits were performed. In these panels, the radial scale
spans 0.5\arcsec--15\arcmin, which is close to \rtwentyfive. In the lower panels, the radial range corresponds to 
$\sim$0.5\arcsec--7\arcmin, which is close to \rfive. The simple entropy fits generally 
do a good job at recovering the true profiles, although at the largest scales in object a, the
temperature is  slightly over-estimated. This most likely reflects partial degeneracy with the 
background. Given its more restrictive assumptions, the \model\ model does a poorer job at 
recovering the true temperature and density profiles, especially in the centres of each system.
Nevertheless, the average properties of the gas, in particular at large scales, are 
recovered reasonably well, which explains why the global properties of the systems are 
measured accurately
(Fig~\ref{fig_compare_derived}).} \label{fig_kt_rho_rxj} \label{fig_kt_rho_n720}
\end{figure*}
To illustrate the performance of the model, we fitted simulated \chandra\ data for two
objects, spanning almost two orders of magnitude in mass. These objects, henceforth termed
objects a and b,  are tailored to 
resemble, respectively, the isolated, Milky Way-mass elliptical galaxy \src\ and the fossil 
group/ cluster \srctwo\ \citep{humphrey11a,humphrey12a}. These systems were chosen
as we have previously studied them over a wide range of physical
scales using the more traditional
approach of spatially resolved spectroscopy. Starting with the published
models for each real system, and assuming \zfe=0.5 times Solar, 
we simulated artificial 
events files corresponding to a 100~ks exposure with \chandra\ ACIS-S3. 
These files were generated as described in \S~\ref{sect_fit_beta} 
\citep[see also][]{humphrey12c}. To represent better real data,
we included an additional component to represent the unresolved
X-ray binary contribution to each system. Specifically, we assumed the combined 
light from the LMXBs was distributed as  de Vaucouleurs profiles, with 
effective radii 25.2\arcsec\ and 6.4\arcsec, respectively, and a 7.3~keV
bremsstrahlung spectrum.
We added a background component with a 
constant surface brightness profile and a plausible spectrum (two soft
APEC thermal plasma components, and a powerlaw, normalized to match
the sky background in the vicinity of \src). This is somewhat idealized as it 
omits the non X-ray background that is important for \chandra\ data at high
energies. For simplicity,
we also ignored the spatial dependence of the effective area and spectral response,
and ignored the spatial point spread of the telescope.

We extracted surface brightness profiles in three energy bands,
0.5--0.9~keV, 0.9--2.0~keV and 2.0--7.0~keV, corresponding to those 
shown in Fig~\ref{fig_hardness}, which were fitted simultaneously.
The model comprised hot gas, a background component and a deVaucouleurs
component to model the unresolved point sources.
The hot gas model was derived by numerically solving Eqns~\ref{eqn_zeta}--\ref{eqn_mg},
assuming $\overline{M_{eff}} \equiv 0$ and $f_{nth}=0$ (\ie\ no nonthermal
pressure), and modelling $\overline{M_{ng}}$ as the sum of a stellar light model
\citep[as discussed in][]{humphrey11a,humphrey12a}, an
NFW dark matter halo and a central black hole. The stellar M/L ratio and the mass and 
concentration of the NFW model were allowed to vary in the fit, and the black hole mass
was fixed at $3\times 10^8$\msun\ for \src\ and $2.4\times 10^9$\msun\ for \srctwo. 
The model assumed a flat abundance profile, but we allowed the overall abundance to fit freely,
fixing the abundance ratios with respect to Fe to their Solar values.

To compute the surface brightness, we evaluated the gas emissivity
at a series of logarithmically spaced radial points
(from $\overline{r}_0$ to 1--2 times the virial radius), and projected it 
using a cubic spline approximation and Eqn~\ref{eqn_projection}. 
To enable direct comparison with our previous work, parameter space exploration was 
carried out with a Bayesian Monte Carlo code 
\citep[version 2.7 of the MultiNest code\footnote{\href{http://www.mrao.cam.ac.uk/software/multinest/}{http://www.mrao.cam.ac.uk/software/multinest/}}:][]{feroz09a}. 
We adopted the appropriate Poisson likelihood function for a set of 
independent data points, and used uniform priors for all variable parameters, similar
to our default analysis in \citet{humphrey11a,humphrey12a}.

The model gave a good fit to both datasets (see Fig~\ref{fig_rxj1159_sb}), and the
abundance was measured reasonably well for each system (\zfe=$0.58\pm0.18$, $0.40\pm0.06$)
In Fig~\ref{fig_compare_derived}, we compare various derived physical parameters obtained from
our fits to the known, true value, demonstrating good overall agreement, and indicating 
that the hydrostatic, narrow band photometric model provides a viable means for constraining 
the global properties of elliptical galaxies, groups and clusters. The constraints are comparable
to those obtained from our fits to the full spatially resolved spectroscopic data of \src\ and
\srctwo\ \citep{humphrey11a,humphrey12a}; for example we measured \mfive\ and \cfive\
in \srctwo\ to 0.05~dex and 6\%, respectively, with both the surface brightness fitting
approach  and the full spectral analysis, if restricted only to the \chandra\ data \citep{humphrey12a}.
In Fig~\ref{fig_kt_rho_rxj}, we show the 
derived temperature and density profiles for each object, which agree well with the 
true profile. For object a, the temperature at large scales is slightly over-estimated,
albeit with large uncertainty. At these scales, the background is comparable to, or exceeds,
the hot gas flux, at least at energies \gtsim 1~keV, which limits the ability to constrain the 
temperature precisely from the hardness ratios. Nevertheless, the global properties of the 
system are very well recovered.  This illustrates the 
potential of narrow band photometry for mass measurements over a range of 
mass scales.

\section{The Scaled Adiabatic Model (\model)} \label{sect_entropy_scaling} \label{sect_scam}
While Eqn~\ref{eqn_powerlaw} is appealing in its
simplicity and has been
used successfully to model the X-ray emission from galaxies and groups
\citep[\eg][]{humphrey08a,humphrey09d,humphrey11a,humphrey12a,humphrey12b}, it
has significant drawbacks. In particular, extrapolating the model to 
large radii involves ad hoc assumptions in a region where the data are 
often sparse, at best, or missing altogether. Furthermore, as the entropy profile
shape becomes more and more complex, additional adjustable parameters are
required, resulting in a large number of potentially
degenerate parameters. These problems are amplified if the S/N of the data is low.

Instead, we propose here to employ an alternative parametrization for the 
entropy distribution. Outside the very central regions, gravity-only models of structure
formation predict a ``baseline'' distribution, 
$\overline{S}\simeq \overline{r}^{1.1}$ \citep{voit05b}, which is 
distorted by non-gravitational heating, especially in the inner
parts of low-mass systems \citep[\eg][]{sanderson03a,cavagnolo09a,pratt10a}. 
Nevertheless, the principal effect of the entropy injection is to lower the 
gas density, rather than raise the temperature \citep[\eg][]{ponman99a,pratt10a,mathews11a},
which is supported by  the universality of cluster and 
group temperature profiles \citep[\eg][]{vikhlinin05a,gastaldello07a}.
Thus, to zeroth order, $\overline{S}\simeq \overline{r}^{1.1} \left( f_g/ f_{b,U}\right)^{-\frac{2}{3}}$, where $f_g$ is the gas fraction. This scaling has been found 
to give a much better match to the observed entropy profiles than the 
baseline model when the enclosed gas fraction, 
$f_g=\overline{M_g}/(\overline{M_g}+\overline{M_{ng}})$ is employed
 \citep{pratt10a,humphrey11a,humphrey12a,humphrey12b}. As we show
below, this model fails to match the data successfully in lower mass
objects, and in the central
parts of all systems (roughly, within the cooling radius). We therefore
begin by refining the parametrization of this scaling relation.

\subsection{Calibrating the model}
\begin{table*}
\begin{tabular}{lllllrrr}\hline
Object & z & \rfive & $R_{max}$ &  $K_{500}$ &  Dataset & Exposure & ref\\
       &   &  (kpc) &  (kpc) & ($keV\ cm^{2}$)  & & ks & \\ \hline
NGC\thin 4125 & 0.004523  & $166\pm18$ & 47 & $20\pm4$  & 2071(C) & 63 & 1\\ 
NGC\thin 720 & 	0.005821 & $199\pm 9$ & 74 & $28\pm 3$ & 7062, 7372, 8448, 8449 (C) & 99 & 2\\
             &           &             &    & & 800009010 (S) & 177 & 2 \\
NGC\thin 6482 & 0.013129  & $205\pm17$ & 97 & $32\pm5$  & 3218(C) & 17 & 3\\
NGC\thin 4261 & 0.007465  & $447\pm45$ & 190 &  $142\pm27$ & 834, 9569(C) & 135 & 4\\
MKW4          & 0.02      & $511\pm24$ & 130 &  $182\pm17$ & 3234(C) & 30 & 1 \\
RXJ\thin 1159+5531 &0.081 & $577\pm17$& 1200 &  $232\pm14$& 4964(C) & 75 & 5\\
                   &      &           &      &            & 804051010(S) & 85 & 5\\
A\thin 1991 & 0.0587 & $718\pm14$ & 680 & $336\pm13$ & 3193(C) & 38 & 6 \\
A\thin 2589 & 0.0414 & $997\pm58$ & 720 & $645\pm77$ & 7190(C) & 53  & 6\\
A\thin 133 & 0.0566 & $1150\pm25$ & 710 & $858\pm26$ & 9897(C) & 69  & 6\\
A\thin 907 & 0.1527 & $1180\pm30$ & 1400 & $966\pm51$ & 3205(C) & 41 & 6\\
A\thin 1413 & 0.1427 & $1290\pm20$ & 1990 & $1140\pm40$ & 5003(C) & 75 & 6\\
\hline
\end{tabular}
\caption{Data used for calibrating the scaled adiabatic model. For each target,
we list the name, redshift (z), \rfive\ derived from our own  fits to the data, the 
maximum radius to which the data were fitted ($R_{max}$), the characteristic 
entropy ($K_{500}$) derived from our fits, the dataset observation identifier for
both the  \chandra\ (C) or \suzaku\ (S) data,
the cleaned exposure time and a reference to the literature describing the data
reduction and analysis. 
References: (1) \citet{humphrey10a}; (2) \citet{humphrey11a}; (3) \citet{humphrey06a}; (4) \citet{humphrey09d}; (5) \citet{humphrey12a}; (6) W.~Liu et al, in preparation.} \label{table_sample}
\end{table*}
\begin{figure*}
\includegraphics[width=7in]{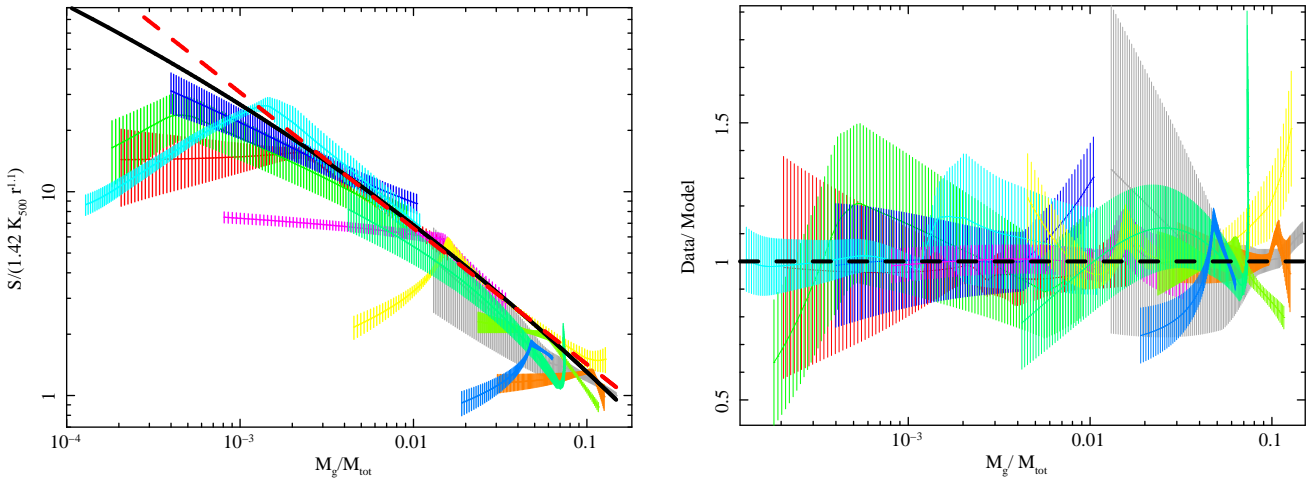}
\caption{Relation between the scaled entropy and the gas fraction for the sample 
of galaxies, groups and clusters (left panel). 
The shaded regions correspond to 1-$\sigma$ confidence intervals.
The solid black line is the best-fitting
simultaneous model fit 
to the data {\em excluding the cool cores of each system} (Eqn~\ref{eqn_scaled1}--\ref{eqn_scaled2}) , which are 
generally apparent as a positive correlation between the enclosed gas fraction and the scaled
entropy. The dashed red line is the expectation if the scaling adopted by \citet{pratt10a} 
is exact. In the right panel, we show the ratio between the observed profiles and the 
\model\ parametrization of the entropy--\fgas\ relation (\S~\ref{sect_scam};
Eqn~\ref{eqn_scaled1}--\ref{eqn_scaled3}). While 
not exact in any given system, we find that \model\ captures the overall shape of the
entropy distribution over a wide range of gas fractions (hence, wide radial range),
including in the cool cores.\label{fig_entropy_scaling}}
\end{figure*}
To calibrate the relation between the gas entropy and the gas fraction, we assembled a 
sample of relaxed, X-ray luminous galaxies, groups and clusters with good quality \chandra\
data to a significant fraction of \rfive\ (Table~\ref{table_sample}). The \chandra\ data were reduced and analyzed to provide
projected temperature and density profiles (deprojected for NGC\thin 4125 and MKW4), as 
described in \citet{humphrey10a,humphrey11a,humphrey12a,liu12c}. For \src\ and \srctwo, we used 
both the \chandra\ and \suzaku\ data described in 
\citet{humphrey11a,humphrey12a}. Using the approach outlined in \S~\ref{sect_entropy_model},
we solved Eqn~\ref{eqn_zeta}--\ref{eqn_mg} to obtain self-consistent models for the deprojected
temperature and density profiles, given parametrized input models for the gravitating 
mass and entropy (for which we used a multiply broken powerlaw, \ie\ Eqn~\ref{eqn_powerlaw}). 
These models were projected along the line of sight, as needed, following the procedure 
described in \citet{humphrey11a}, and fitted to the observed data.

While this approach limits the flexibility of the inferred entropy profiles by imposing a 
parametric form, it has the advantage of providing complementary entropy, mass and 
gas fraction profiles. Furthermore, models of this form have previously been shown
to provide an adequate fit to galaxies, groups and clusters. In Fig~\ref{fig_entropy_scaling},
we show the scaled entropy as a function of enclosed gas fraction.
The data are shown for each system, starting at the smallest resolvable scales
and extending to $\sim$\rfive. At large \fg\ values, the scaled entropy profiles
show little scatter, indicating scaling similar to that pointed out by \citet{pratt10a}.
However,
we note that the \citet{pratt10a} scaling does not exactly describe the relationship,
as it deviates at smaller values of \fg. 
At the smallest \fg\ values, approximately corresponding to the cooling core, the 
relationship becomes non-monotonic, and exhibits significant scatter from system to
system. 

To parametrize the relationship, we begin by assuming a relation of the form:
\begin{eqnarray}
\overline{S}=s_0 +(1-s_0) \overline{r}^{1.1} \overline{S}^\prime\left( \frac{f_g}{f_{b,U}}\right) \label{eqn_scaled1}
\end{eqnarray}
where $s_0$ is a central entropy value, and $\overline{S}^\prime$ is an empirical scaling
function that is chosen to fit observations. With experimentation, we found that 
we were able to fit the data satisfactorily (Fig~\ref{fig_entropy_scaling}, right panel) 
by employing a model of the form:
\begin{eqnarray}
\overline{S}^\prime & = & s_n \exp\left( -s_e \left( \frac{f_g}{f_{b,U}}\right)^{s_i} \right)  \label{eqn_scaled2}
\end{eqnarray}
if $f_g\ge f_{g1}$, or 
\begin{eqnarray}
\overline{S}^\prime & = & s_n  \exp\left( -s_e \left( \frac{f_{g1}}{f_{b,U}}\right)^{s_i} \right) \left( \frac{f_g}{f_{g1}} \right)^{-s_a} \label{eqn_scaled3}
\end{eqnarray}
otherwise, where $s_n$, $s_e$, $s_i$, $f_{g1}$, $s_0$ and $s_a$ are adjustable parameters. In practice,
we found that $s_n = 3440$, $s_e= 10.6$ and $s_i = 0.076$ gave a 
good description of the data in most cases.
The remaining parameters, which all relate to the distribution of the entropy 
within the core of the system, need to be adjusted from object to object.

\subsection{Performance of the model}
\begin{table*}
\begin{tabular}{ll}\hline
Parameter & Recommended value \\ \hline
$\log_{10}$ \mfive & free parameter\\
$\log_{10}$ \cfive & free parameter\\
M/L$_*$ & free parameter  \\ 
$M_{BH}$ & free parameter (where possible) \\
$s_0$ & free parameter \\
$s_a$ & free parameter \\
$f_{g1}$ & free parameter \\
$s_n$ & 3440 \\
$s_e$ & 10.6 \\
$s_i$ & 0.076 \\ 
$\overline{\zeta}$ & free parameter \\
\zfe & free parameter\\
\hline
\end{tabular}
\caption{Summary of variable and fixed parameters for the \model\ model.}\label{table_parameters}
\end{table*}
Having eliminated $\overline{S}$ by application of Eqn~\ref{eqn_scaled1}--\ref{eqn_scaled3},
we solved Eqn~\ref{eqn_zeta}--\ref{eqn_mg} and computed surface brightness profiles,
which we fitted to the data for \src\ and \srctwo, exactly as described in 
\S~\ref{sect_simple_entropy_performance}. In Table~\ref{table_parameters} we summarize the 
fitted and fixed parameters.
In Fig~\ref{fig_compare_derived} we compare the derived properties for the simulated 
objects a and b with the true values, finding overall excellent agreement. The gas abundance was
measured accurately (\zfe=$0.58\pm0.18$ and $0.59\pm0.06$, respectively)
In Fig~\ref{fig_kt_rho_rxj}, we show the derived temperature and density profiles for 
both systems. Overall, the fits are reasonable, but in the inner regions, we see 
statistically significant (albeit not dramatic) deviations of the best fitting model from the true profiles.
Nevertheless, at large scales the \model\ model does a good job of capturing the 
shapes of the profiles, explaining why the global parameters are so well recovered. 
 These results
give us confidence in the utility of the Scaled Adiabatic Model (\model) to derive
global physical parameters from narrow-band photometry.

Since the \model\ model imposes restrictions on the entropy profile distribution at large
scales, it is ideally suited for fitting low-to-modest S/N data. To demonstrate its
usefulness, we simulated data corresponding to a 5~ks snapshot of object a, corresponding
to only 5\%\ of the exposure used in our previous discussion. Fitting the 
surface brightness profiles, we found that the global 
parameters were accurately recovered. While the precision of the constraints was 
poorer than when fitting the same model to the 100~ks exposure, 
the global parameters were, nevertheless, recovered to within $\sim$0.1~dex or better.

\section{Discussion} \label{sect_discussion}
\subsection{Narrow band photometry as a practical tool}
We have demonstrated that narrow band photometry provides an effective alternative to
spatially resolved spectroscopy for determining detailed mass profiles of 
X-ray luminous galaxies, groups and clusters. Since surface photometry is, in general, 
less demanding of data than spectroscopy,
the narrow band photometric method presented here is ideally suited to systems with
modest S/N, for which spatially resolved spectroscopy is not realistic. Without the 
intermediate step of measuring
temperature and density profiles (from single phase plasma model fits), biases 
due to the projection of different temperature gas components along the line of sight are 
effectively minimized, and the difficulty of accounting for covariance between the 
temperature and density data points downstream is circumvented.

In our fully Bayesian forward-fitting analysis, a model for the 
surface brightness profile in each band is constructed based on a physical model for the 
mass distribution and the entropy profile, assuming hydrostatic equilibrium, a single
phase gas and spherical symmetry, and is directly fitted to the data. 
Although real galaxy
clusters are not exactly spherical, on average the spherical approximation does not 
introduce significant bias into their inferred global properties, such as the virial
mass, halo concentration, and gas fraction \citep{buote11c}. Similarly, the hydrostatic 
approximation, provided the system is sufficiently relaxed in its X-ray morphology, is 
likely to be accurate at least to the $\sim$25\%\ level \citep[][and references therein]{buote11a}.
In regions where the gas is morphologically relaxed, spatially resolved spectroscopy 
is generally consistent with a single phase plasma, provided the temperature gradient is 
not strong across the aperture \citep[\eg][]{molendi01b,lewis02a,humphrey05a}.
Similarly, the range of temperatures in the cores of clusters measured within the \xmm\ RGS aperture 
\citep{peterson03a} is broadly  consistent with the observed central temperature gradients. 
Limited multiphase gas has, nevertheless, been inferred in regions of strong interaction
between the hot gas and ejecta from a central AGN \citep[\eg][]{molendi02a,buote03a,david11a}.
Recent work has also invoked multiphase gas as a possible explanation 
of the unexpected flattening of cluster entropy profiles outside $\sim$\rfive\ \citep{simionescu11a,urban11a},
although observations of relaxed clusters suggest such behaviour (and, hence, the postulated
clumping) is not ubiquitous \citep{humphrey12a,miller11a,eckert12a}.

While it is relatively straightforward to identify and exclude data from the 
disturbed cores of clusters (where the hydrostatic and single phase approximations
may break down), the effects of possible clumping at
large scales are more troubling, due to projection effects. Still, if one focuses
on regions well within projected \rfive, the projected emission from outside
\rfive\ is unlikely to contribute significantly along the line of sight, in which case
the results should not depend sensitively on the behaviour of the entropy in the 
outskirts of the cluster. This can be tested explicitly by varying the maximum radius
used during the projection calculation. Application of the photometry method to the outskirts of 
clusters can, of course, also be used to test the need for clumping or deviations from
hydrostatic equilibrium in those regions. If necessary, given a physical model 
for the clumping, the expression for $\epsilon_{\nu^\prime}$  can be modified to 
compensate and allow a useful measurement of the gravitating mass even in those regions.

While the number of energy bands used is entirely arbitrary, we found that 
three standard bands (0.5--0.9~keV, 0.9--2.0~keV and  2.0--7.0~keV) were sufficient 
to constrain the gas temperature and total abundance from \chandra\ data over 
a fairly wide range of parameter space. Limiting the number of photometric bands
is desirable, since it minimizes the number of adjustable parameters needed to account for
the background. Nevertheless, if the data permit, the number of energy bands can be
expanded to focus on regions of particular interest that may help constrain the 
relative abundance of other species, such as O, Ne, Mg and Si.

\subsection{The entropy parametrization}
We present two classes of hydrostatic model that enable the gravitating mass to be
inferred. Both involve a physical model for the (non-gas) gravitating mass
(comprising dark matter halo, stars and central black hole), and solving the hydrostatic
equations (Eqn~\ref{eqn_zeta}--\ref{eqn_mg}) by imposing a constraint on the entropy distribution. 
The performance of the former approach, which entails adopting an arbitrary parametrization
for the entropy profile, has been demonstrated in our previous, spatially resolved
spectrocopic fits to X-ray data \citep{humphrey08a,humphrey09d,humphrey11a,humphrey12a,humphrey12b}, and, 
independently, in fits to Sunyaev-Zeldovich data \citep{allison11a}. Combined with
narrow-band photometry, we have demonstrated that precise, accurate constraints can be 
obtained on the global properties of objects over a wide mass range.

The latter entropy parametrization is more suitable for lower S/N data, since it involves exploiting
an empirical scaling relation between the gas fraction and entropy profile outside the
cooling core, effectively eliminating the arbitrariness in the extrapolation of the 
entropy profile to large scales. This scaling relation is unlikely to be exact in any given
system, which introduces systematic errors into the recovered temperature and density profiles,
especially in the system cores. However, with the extra constraints at large scale, we found
that the reduction in the error-space did not give rise to appreciable biases in the derived
global system properties. The utility of this method is illustrated by the fact that, even for 
a shallow, snapshot observation of a realistic low-mass object, we were still able to recover
the global parameters to within $\sim$0.1~dex.

\subsection{Abundance gradients} \label{sect_discuss_abundance}
For simplicity, our simulations were for systems with a constant metallicity profile. In reality 
morphologically relaxed systems tend to have centrally-peaked metallicity distributions
\citep[\eg][]{degrandi01a,buote03b,humphrey05a,johnson11a} and so, if the properties in 
the centre of the system are important it is necessary to modify the model to incorporate 
such a gradient. As there is no {\em de facto} standard model for such purposes, we
recommend a simple ansatz, such as:
\begin{equation}
Z_{Fe} = Z_{Fe,1}+\left( Z_{Fe,0}-Z_{Fe,1}\right) \left( 1+ \frac{r^2}{r^2_{Fe}}\right)^{-\alpha_{Fe}}
\label{eqn_abund_gradient}
\end{equation}
where $Z_{Fe,0}$ and $Z_{Fe,1}$, $r_{Fe}$ and $\alpha_{Fe}$ are 
adjustable fit parameters, and we initially assume that the ratio of the abundance of the other 
metals to Fe does not vary with radius. To verify that adopting such a parametrization does
not appreciably degrade the constraints on the global system parameters, we re-simulated 
object b, assuming Eqn~\ref{eqn_abund_gradient}. We set $log_{10} Z_{Fe,0}=0.0$, $log_{10} Z_{Fe,1}=-0.5$, 
$log_{10} r_{Fe}=1.0$~[kpc] and $\alpha_{Fe}=0.5$. We fitted the surface brightness profiles, using the 
\model\ model (this time incorporating Eqn~\ref{eqn_abund_gradient}) 
and allowing the parameters $Z_{Fe,0}$, $r_{Fe}$ and $\alpha_{Fe}$ to be fitted. We summarize the 
constraints on the global parameters in Fig~\ref{fig_compare_derived}, demonstrating that 
incorporating an abundance gradient into our fits does not significantly degrade the 
level of constraints we can obtain. Furthermore, with this approach, we were also able to constrain the 
abundance profile reliably (we measured $log_{10} Z_{Fe,0}=0.06\pm0.09$, $log_{10} r_{Fe}=0.97^{+0.20}_{-0.15}$ [kpc] 
and $\alpha_{Fe}=1.14^{+0.26}_{-0.40}$).

\subsection{The Sunyaev-Zeldovich effect}
While we have focused on the use of the \model\ model in the context of narrow-band X-ray 
photometry, it actually has broader implications. In particular, it is a potentially powerful
tool for constraining the global properties of galaxy clusters from the 
Sunyaev-Zeldovich (S-Z) effect \citep[][for a review]{birkinshaw99a}. The key observable is
the Compton y-parameter,
\begin{equation}
y = \frac{\sigma_T}{m_e c^2} \int_{los} n_e kT dl
\end{equation}
where $\sigma_T$ is the Thompson scattering cross-section, $m_e$ is the mass of the electron,
$c$ is the speed of light, $n_e$ is the electron number density, and the integral is along
the line-of-sight (los). In comparison to X-ray emission, for which the emissivity depends on
the density {\em squared}, the y-parameter is more sensitive to the properties of the gas at large 
radii. 

In order to derive the physical properties of a system from the inferred y-parameter, it is 
necessary to construct a model for the spatial variation in both density and temperature. 
Models employing {\em ad hoc} parametrization for the thermodynamical state of the gas,
for example the isothermal $\beta$-model, or even the simple entropy-based model discussed
in \S~\ref{sect_entropy_model} \citep[for the application of this kind of model to S-Z data, 
see][]{allison11a}, cannot be guaranteed to capture the true behaviour of the gas when 
extrapolated to large scales. On the other hand, the thermodynamic behaviour of the gas at large
radii predicted by the \model\ model is determined entirely by the underlying scaling 
relation, and it is clear from Fig~\ref{fig_kt_rho_rxj} that it is sufficiently accurate to model 
morphologically relaxed objects with very different mass distributions, density and temperature 
profiles. While it remains to be seen whether there exist clusters for which the 
empirical \model\ scaling does not (at least approximately) hold, the well-determined 
asymptotic behaviour of the model is very desirable in S-Z studies, and so we strongly 
advocate its use.

\section*{Acknowledgments}
We are grateful to
Christina Topchyan for helping extensively test the models.
We thank Wenhao Liu for providing his temperature and density profiles. 
PJH and DAB acknowledge partial support from NASA under Grant NNX10AD07G, 
issued through the office of Space Science Astrophysics Data Program. PJH also
acknowledges support through a Gary McCue Fellowship, offered through UC
Irvine.

\appendix
\section{Computing the X-ray Surface Brightness Profile} \label{sect_SB}
\begin{figure}
\includegraphics[width=3.5in]{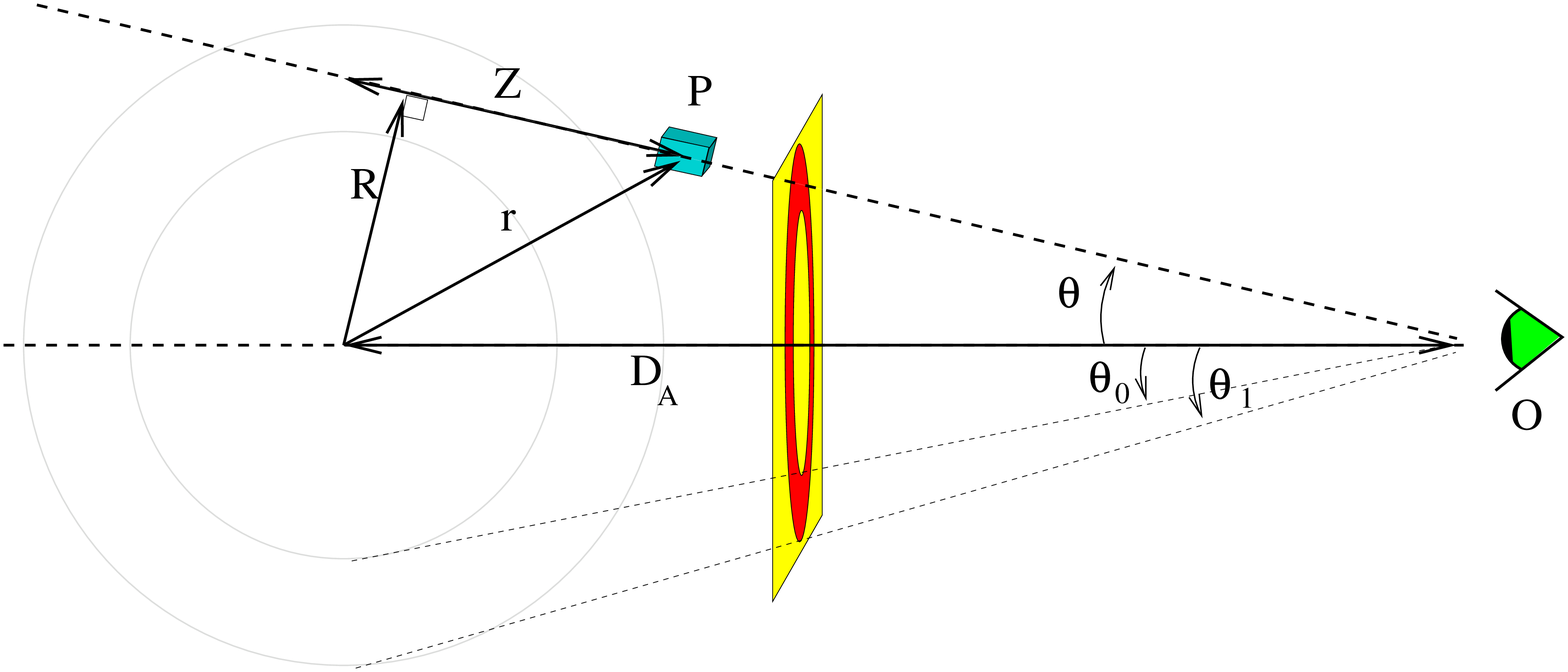}
\caption{Schematic view showing the projection of the object onto the plane of the 
sky (shown schematically in yellow). The gas properties at point P are a function only
of the radial distance from the centre of the object (r), while the distance to
the observer (O) is $D_A$. The distance along the line of sight to the parcel is 
denoted as $Z$, and the inner and outer radii of an arbitrary shell of gas (shown in 
red) correspond to incidence angles $\theta_0$ and $\theta_1$, respectively. 
\label{fig_projection}}
\end{figure}
In this section, we discuss the efficient computation of the 
X-ray surface brightness model, which is critical for making narrow-band
photometry a useful tool for model fitting.

\subsection{The Surface Brightness Distribution}
Consider a small parcel of gas of volume dV, which is emitting a total
(rest frame) photon flux (\ie\ photons s$^{-1}$) of $\epsilon_{\nu^\prime} dV d\nu^\prime$ at frequency $\nu^\prime$. Let the system be spherically symmetric (see Fig~\ref{fig_projection}), so that 
$\epsilon_{\nu^\prime}$ is a function of distance, r, from the centre
of the system. The total photon flux incident on a telescope aperture of area A 
placed at angular size distance $D_A$ from the gas parcel will be:
\begin{equation}
N_\nu d\nu= \frac{A \epsilon_{\nu^\prime}(r) dV d\nu^\prime}{4\pi D_A^2 (1+z)^3} = 
\frac{A \epsilon_{\nu^\prime}(r) d\Omega dZ d\nu^\prime}{4\pi (1+z)^3} \label{eqn_dN}
\end{equation}
where $z$ is the observed redshift of the gas parcel, 
$\nu=\nu^\prime (1+z)^{-1}$ is the {\em observed} photon frequency, 
$d\Omega$ is the angular size subtended by the gas parcel at the telescope,
and $Z$ is the line of sight coordinate in the rest frame of the 
gas parcel, so that $r=r(\theta,Z)$.

The arriving photons pass through the telescope optics and are (possibly)
detected in a focal plane detector. The detected
photon flux at position (x,y) in ``pulse height channel'' (\ie\ energy
bin) $i$ is given by:
\begin{eqnarray}
dC_i(x,y) = N_\nu d\nu\ P_\nu(x,y,\theta,\phi) E_\nu(x,y,\theta,\phi) \mathbb{R}_i(x,y,\nu)
\end{eqnarray}
where $P_\nu (x,y,\theta,\phi)$ codifies the telescope spatial point 
spread function at frequency $\nu$ for photons arriving with incident
angle $\theta$ and azimuthal angle $\phi$ (\ie\ it represents the 
likelihood that such a photon would be detected at position (x,y)), 
$E_\nu$ is the combined
detection efficiency of the optics and detector, and $\mathbb{R}_i$ codifies
the spectral redistribution function, \ie\ indicating the likelihood
that a photon with frequency $\nu$ that is detected at position (x,y) 
will be assigned to pulse height channel i.

To make further progress, we divide the sky into a series of concentric
annuli, the centres of which coincide with the projected centre of the 
object. Annulus $j$ has inner and outer radii $\theta_j$ and 
$\theta_{j+1}$, respectively. Now, let us compute the total counts $C_i^{k}$
detected in pulse height channel $i$ within some (arbitrary) region of the 
detector, defined by the function $\delta_k(x,y)$, which is 1 in
the region, or 0 elsewhere. It follows that:
\begin{eqnarray}
C_i^{k}  =   \int \delta_k(x,y) dx\ dy\ \sum_j \int_0^{2\pi} d\phi \int_{\theta_j}^{\theta_{j+1}} \sin \theta d\theta\  \int_{-\infty}^{\infty} dZ \bigg[\nonumber \\ \int d\nu^\prime\frac{A \epsilon_{\nu^\prime}(r)}{4\pi(1+z)^3} P_\nu(x,y,\theta,\phi) E_\nu(x,y,\theta,\phi)  \mathbb{R}_i(x,y,\nu) \bigg]
\end{eqnarray}
where x and y are coordinates on the detector.
Now, let annulus $j$ be sufficiently narrow that $P_\nu(x,y,\theta,\phi) \simeq P_\nu(x,y,\theta_j,\phi)$
and $E_\nu(x,y,\theta,\phi) \simeq E_\nu(x,y,\theta_j,\phi)$, so that we can
rewrite
\begin{eqnarray}
C_i^{k} \simeq \sum_j \int d\nu^\prime \bigg[ \int \delta_k(x,y) dx\ dy\ \mathbb{R}_i(x,y,\nu)  \nonumber\\ \int_0^{2\pi} d\phi  P_\nu(x,y,\theta_j,\phi)
A\ E_\nu(x,y,\theta_j,\phi)\bigg] \nonumber \\ \times \bigg[ \int_{\theta_j}^{\theta_{j+1}}\sin\theta d\theta \int_{-\infty}^{\infty} \frac{\epsilon_{\nu^\prime}(r)}{4\pi(1+z)^3}dZ \bigg]
\end{eqnarray}
It is convenient to rewrite this as:
\begin{eqnarray}
C_i^{k} & \simeq&  \sum_j \int d\nu^\prime \bigg[ \overline{\mathbb{R}}^k_i(\nu) \overline{P}_{\nu}^{j\rightarrow k} A_{eff}^{j\rightarrow k} \bigg] \times \bigg[ n_\nu^j \bigg]
\end{eqnarray}
where the quantity $R^k_i(\nu)$ is the spectral redistribution function, $P_{\nu}^{j\rightarrow k}$ is the 
spatial ``point response function'' linking sky annulus $j$ to detector region $k$,
$A_{eff}^{j\rightarrow k}$ is the ``effective area'' linking annulus $j$ to detector region $k$, and 
\begin{eqnarray}
& & n_\nu^j=  \int_{\theta_j}^{\theta_{j+1}}\sin\theta d\theta \int_{-\infty}^{\infty} \frac{\epsilon_{\nu^\prime}(r)}{4\pi(1+z)^3}dZ \label{eqn_n_nu_j1} \\ 
& & =  \int_{\theta_j}^{\theta_{j+1}} \sin \theta d\theta \int_{D_A \sin \theta}^{\infty}
\frac{2}{(1+z)^3} \frac{r ( \epsilon_{\nu^\prime} / 4\pi ) dr}{\sqrt{r^2-D_A^2\sin^2\theta}}  \label{eqn_n_nu_j}
\end{eqnarray}
Changing the order of integration, this becomes:
\begin{eqnarray}
n^j_\nu & = & \frac{2}{(1+z)^3}\int_{R_j}^\infty \frac{\epsilon_{\nu^\prime}}{4 \pi} r\ dr \int_{\theta_j}^{\Theta_{j+1}} \frac{\sin \theta d\theta}{\sqrt{r^2-D_A^2 \sin^2\theta}} \\
 & \equiv & \frac{2}{(1+z)^3}\int_{R_j}^\infty \frac{\epsilon_{\nu^\prime}}{4\pi} r\  I_j(r) dr
\end{eqnarray}
where $\Theta_{j+1}={\rm min}\left({\rm asin}(r/D_A), \theta_{j+1}\right)$ is introduced due to the 
change of the order of integration and the requirement that $r>D_A sin \theta$, and
$R_j=D_A \sin \theta_j$, and $I_j(r)=\int_{\theta_j}^{\Theta_{j+1}} \sin \theta d\theta/\sqrt{r^2-D_A^2 \sin^2\theta} \simeq \left( \sqrt{r^2-D_A^2\theta_j^2} - \sqrt{r^2-D_A^2\Theta_{j+1}^2}\right)/D_A^2$, applying the small angle approximation $\sin \theta \simeq \theta$ and integrating.

Now, let us define the X-ray surface brightness in a particular PHA channel band ($i_0$--$i_1$),
\begin{eqnarray}
SB^k_{i_0,i_1} & = & \frac{\sum_{i=i_0}^{i_1} C_i^k}{A_k} = \frac{\sum_j \int d\nu^\prime \overline{P}_\nu^{j\rightarrow k} A_{eff}^{j\rightarrow k} n_\nu^j \sum_{i=i_0}^{i_1} \overline{\mathbb{R}}_i^k}{A_k} \nonumber
\end{eqnarray}
where $A_k\equiv\int \delta_k(x,y) dx\ dy$ is the area of detector region $k$.
Provided the PHA channel range $i_0$--$i_1$ is sufficiently narrow, and the redistribution function
$\overline{\mathbb{R}}^k_i(\nu)$ is not too broad, it is often sufficient to approximate
$\overline{P}^{j\rightarrow k}_\nu \simeq \overline{P}^{j\rightarrow k}_{\nu_{0,1}}$, where
$\nu_{0,1}$ is the mean frequency of photons that would be detected in PHA bins $i_0$--$i_1$.
Substituting in for $n_\nu^j$ from Eqn~\ref{eqn_n_nu_j}, we get:
\begin{equation}
SB^k_{i_0,i_1} \simeq \frac{\sum_j \overline{P}^{j\rightarrow k}_{\nu_{0,1}} \int_{R_j}^\infty r I_j dr \int d\nu^\prime A_{eff}^{j\rightarrow k} \epsilon_{\nu^\prime} \sum_{i=i_0}^{i_1} \overline{\mathbb{R}}^k_i}{A_k(1+z)^3}  \label{eqn_sigma}
\end{equation}

\subsection{Piecewise polynomial approximation}
To evaluate  Eqn~\ref{eqn_sigma} efficiently, it is convenient to approximate the $\nu$ integral
as a piecewise polynomial function of r (for example, a cubic spline function), \ie\
\begin{eqnarray}
\int d\nu^\prime A_{eff}^{j\rightarrow k} \epsilon_{\nu^\prime}(r) \sum_{i=i_0}^{i_1} \overline{\mathbb{R}}^k_i 
\simeq \sum_m \sqcap_m(r) \sum_{n=0} a_{mn}^{j\rightarrow k} r^n
\end{eqnarray}
where $\sqcap_m(r) = 1$ if $r_m\le r< r_{m+1}$, and $0$ otherwise,
$r_m$ are a series of arbitrarily spaced reference points where the integral is 
explicitly evaluated, and $a_{mn}^{j \rightarrow k}$ are a set of coefficients that are 
chosen to ensure that the approximation is exact at each reference point. 
In general, $\epsilon_{\nu^\prime}(r_m)$ will be a function of the gas density and 
temperature at radius $r_m$, which can be computed from Eqn~\ref{eqn_zeta}--\ref{eqn_mg}, 
as well as the 
gas abundance, for which one can adopt an arbitrary parametrization.
In practice, the integral is generally computed approximately at each radius $r_m$ 
(for example, within a spectral fitting package such as \xspec):
\begin{eqnarray}
\int d\nu^\prime \epsilon_{\nu^\prime} \overline{AR}^{j\rightarrow k}_\nu
\simeq \sum_p  \overline{AR}^{j\rightarrow k}_{\overline{\nu_p}} 
\int_{\nu^\prime_p}^{\nu^\prime_{p+1}} d\nu^\prime \epsilon_{\nu^\prime} \nonumber
\end{eqnarray}
where $\overline{AR}^{j\rightarrow k}_\nu=A_{eff}^{j\rightarrow k}(\nu) \sum_{i=i_0}^{i_1} \overline{\mathbb{R}}^k_i(\nu) $, $\nu^\prime_p\equiv (1+z) \nu_p $ are a set of reference frequencies and $\overline{\nu}_{p}$ is an appropriate
averaged frequency between $\nu_p$ and $\nu_{p+1}$.

Hence
\begin{eqnarray}
SB^k_{i_0,i_1} \simeq  \frac{\sum_j \overline{P}^{j\rightarrow k}_{\nu_{0,1}} \sum_{m,n} a_{mn}^{j\rightarrow k} \int_{max(R_j,r_m)}^{max(R_j,r_{m+1})} r^{1+n} dr I_j(r) }{A_k(1+z)^3} \nonumber \\
= \sum_j \frac{\overline{P}^{j\rightarrow k}_{\nu_{0,1}}}{A_k(1+z)^3} \sum_{m,n} a_{mn}^{j\rightarrow k} \bigg[   \int_{R^\uparrow_{j,m}}^{R^\downarrow_{j+1,m+1}}  r^{n+1} \sqrt{r^2-R_j^2}\ dr \nonumber \\  + \int_{R^\uparrow_{j+1,m}}^{R^\uparrow_{j+1,m+1}} r^{n+1} \left( \sqrt{r^2-R_j^2} - \sqrt{r^2-R_{j+1}^2}\right) dr \bigg] \label{eqn_projection}
\end{eqnarray}
where $R^\uparrow_{j,m} = max(R_j,r_m)$ and $R^\downarrow_{j,m} = min(R_j,r_m)$ 

The integrals can be evaluated analytically. If we define
\begin{eqnarray}
J_{n}(r,a)& =& \int_a^r r^n \sqrt{r^2-a^2}\ dr  \label{eqn_Jn}
\end{eqnarray}
we note that
\begin{eqnarray}
J_0(r,a) & = & \frac{r}{2}\sqrt{r^2-a^2} - \frac{a^2}{2} \ln \left(\frac{ r+\sqrt{r^2-a^2}}{a}\right) \nonumber\\
J_1(r,a) & = & \frac{1}{3} \left( r^2-a^2\right)^\frac{3}{2}  \nonumber
\end{eqnarray}
To compute the remaining terms, we integrate Eqn~\ref{eqn_Jn} by parts 
and rearrange it to give the recurrence relation
\begin{eqnarray}
J_n(r,a)& =& \frac{r^{n-1} \left( r^2-a^2\right)^\frac{3}{2} +\left(n-1 \right)a^2 J_{n-2}}{2+n} \label{eqn_Jn_recurrence}
\end{eqnarray}
Thus, the computation of $SB^k_{i_0,i_1}$ reduces to the sum of a  finite series.

\bibliographystyle{apj_hyper}

\begin{thebibliography}{98}
\expandafter\ifx\csname natexlab\endcsname\relax\def\natexlab#1{#1}\fi

\bibitem[{{Abadi} {et~al.}(2010){Abadi}, {Navarro}, {Fardal}, {Babul}, \&
  {Steinmetz}}]{abadi10a}
\href{http://adsabs.harvard.edu/abs/2010MNRAS.407..435A}{{Abadi}, M.~G.,
  {Navarro}, J.~F., {Fardal}, M., {Babul}, A., \& {Steinmetz}, M.} 2010,
  \mnras, 407, 435

\bibitem[{{Allen}(1998)}]{allen98a}
\href{http://adsabs.harvard.edu/cgi-bin/nph-bib_query?bibcode=1998MNRAS.296..392A&db_key=AST}{{Allen},
  S.~W.} 1998, \mnras, 296, 392

\bibitem[{{Allison} {et~al.}(2011){Allison}, {Taylor}, {Jones}, {Rawlings}, \&
  {Kay}}]{allison11a}
\href{http://adsabs.harvard.edu/abs/2011MNRAS.410..341A}{{Allison}, J.~R.,
  {Taylor}, A.~C., {Jones}, M.~E., {Rawlings}, S., \& {Kay}, S.~T.} 2011,
  \mnras, 410, 341

\bibitem[{{Arnaud}(2009)}]{arnaud09a}
\href{http://adsabs.harvard.edu/abs/2009A\%26A...500..103A}{{Arnaud}, M.} 2009,
  \aap, 500, 103

\bibitem[{{Asplund} {et~al.}(2004){Asplund}, {Grevesse}, \&
  {Sauval}}]{asplund04a}
\href{http://arxiv.org/abs/astro-ph/0410214}{{Asplund}, M., {Grevesse}, N., \&
  {Sauval}, J.} 2004, in {Cosmic abundances as records of stellar evolution and
  nucleosynthesis}, ed. F.~N. {Bash} \& T.~G. {Barnes} (ASP Conf. series),
  astro-ph/0410214

\bibitem[{{Binney} {et~al.}(1990){Binney}, {Davies}, \&
  {Illingworth}}]{binney90a}
\href{http://adsabs.harvard.edu/cgi-bin/nph-bib_query?bibcode=1990ApJ...361...78B&db_key=AST}{{Binney},
  J.~J., {Davies}, R.~L., \& {Illingworth}, G.~D.} 1990, \apj, 361, 78

\bibitem[{{Birkinshaw}(1999)}]{birkinshaw99a}
\href{http://adsabs.harvard.edu/abs/1999PhR...310...97B}{{Birkinshaw}, M.}
  1999, \physrep, 310, 97

\bibitem[{{Buote}(2000)}]{buote00a}
\href{http://adsabs.harvard.edu/cgi-bin/nph-bib_query?bibcode=2000MNRAS.311..176B&db_key=AST}{{Buote},
  D.~A.} 2000, \mnras, 311, 176

\bibitem[{{Buote} {et~al.}(2004){Buote}, {Brighenti}, \& {Mathews}}]{buote04c}
\href{http://adsabs.harvard.edu/cgi-bin/nph-bib_query?bibcode=2004ApJ...607L..91B&db_key=AST}{{Buote},
  D.~A., {Brighenti}, F., \& {Mathews}, W.~G.} 2004, \apjl, 607, L91

\bibitem[{{Buote} \& {Fabian}(1998)}]{buote98c}
\href{http://adsabs.harvard.edu/cgi-bin/nph-bib_query?bibcode=1998MNRAS.296..977B&db_key=AST}{{Buote},
  D.~A. \& {Fabian}, A.~C.} 1998, \mnras, 296, 977

\bibitem[{{Buote} {et~al.}(2007){Buote}, {Gastaldello}, {Humphrey},
  {Zappacosta}, {Bullock}, {Brighenti}, \& {Mathews}}]{buote07a}
\href{http://adsabs.harvard.edu/abs/2007ApJ...664..123B}{{Buote}, D.~A.,
  {Gastaldello}, F., {Humphrey}, P.~J., {Zappacosta}, L., {Bullock}, J.~S.,
  {Brighenti}, F., \& {Mathews}, W.~G.} 2007, \apj, 664, 123

\bibitem[{{Buote} \& {Humphrey}(2012{\natexlab{a}})}]{buote11a}
\href{http://adsabs.harvard.edu/abs/2012ASSL..378..235B}{{Buote}, D.~A. \&
  {Humphrey}, P.~J.} 2012{\natexlab{a}}, in Astrophysics and Space Science
  Library, Vol. 378, Astrophysics and Space Science Library, ed. {D.-W.~Kim \&
  S.~Pellegrini}, 235, (arXiv:1104.0012)

\bibitem[{{Buote} \& {Humphrey}(2012{\natexlab{b}})}]{buote11b}
\href{http://adsabs.harvard.edu/abs/2012MNRAS.420.1693B}{{Buote}, D.~A. \&
  {Humphrey}, P.~J.} 2012{\natexlab{b}}, \mnras, 420, 1693

\bibitem[{{Buote} \& {Humphrey}(2012{\natexlab{c}})}]{buote11c}
\href{http://adsabs.harvard.edu/abs/2012MNRAS.421.1399B}{{Buote}, D.~A. \&
  {Humphrey}, P.~J.} 2012{\natexlab{c}}, \mnras, 421, 1399

\bibitem[{{Buote} {et~al.}(2003{\natexlab{a}}){Buote}, {Lewis}, {Brighenti}, \&
  {Mathews}}]{buote03a}
\href{http://adsabs.harvard.edu/cgi-bin/nph-bib_query?bibcode=2003ApJ...594..741B&amp;db_key=AST}{{Buote},
  D.~A., {Lewis}, A.~D., {Brighenti}, F., \& {Mathews}, W.~G.}
  2003{\natexlab{a}}, \apj, 594, 741

\bibitem[{{Buote} {et~al.}(2003{\natexlab{b}}){Buote}, {Lewis}, {Brighenti}, \&
  {Mathews}}]{buote03b}
\href{http://adsabs.harvard.edu/cgi-bin/nph-bib_query?bibcode=2003ApJ...595..151B&amp;db_key=AST}{{Buote},
  D.~A., {Lewis}, A.~D., {Brighenti}, F., \& {Mathews}, W.~G.}
  2003{\natexlab{b}}, \apj, 595, 151

\bibitem[{{Buote} \& {Tsai}(1995)}]{buote95a}
\href{http://adsabs.harvard.edu/cgi-bin/nph-bib_query?bibcode=1995ApJ...439...29B&db_key=AST}{{Buote},
  D.~A. \& {Tsai}, J.~C.} 1995, \apj, 439, 29

\bibitem[{{Cappellari} {et~al.}(2002){Cappellari}, {Verolme}, {van der Marel},
  {Kleijn}, {Illingworth}, {Franx}, {Carollo}, \& {de Zeeuw}}]{cappellari02a}
\href{http://adsabs.harvard.edu/abs/2002ApJ...578..787C}{{Cappellari}, M.,
  {Verolme}, E.~K., {van der Marel}, R.~P., {Kleijn}, G.~A.~V., {Illingworth},
  G.~D., {Franx}, M., {Carollo}, C.~M., \& {de Zeeuw}, P.~T.} 2002, \apj, 578,
  787

\bibitem[{{Cash}(1979)}]{cash79a}
\href{http://adsabs.harvard.edu/abs/1979ApJ...228..939C}{{Cash}, W.} 1979,
  \apj, 228, 939

\bibitem[{{Cavagnolo} {et~al.}(2009){Cavagnolo}, {Donahue}, {Voit}, \&
  {Sun}}]{cavagnolo09a}
\href{http://adsabs.harvard.edu/abs/2009ApJS..182...12C}{{Cavagnolo}, K.~W.,
  {Donahue}, M., {Voit}, G.~M., \& {Sun}, M.} 2009, \apjs, 182, 12

\bibitem[{{Cavaliere} \& {Fusco-Femiano}(1976)}]{cavaliere76a}
\href{http://adsabs.harvard.edu/abs/1976A\%26A....49..137C}{{Cavaliere}, A. \&
  {Fusco-Femiano}, R.} 1976, \aap, 49, 137

\bibitem[{{Cavaliere} \& {Fusco-Femiano}(1978)}]{cavaliere78a}
\href{http://adsabs.harvard.edu/cgi-bin/nph-bib_query?bibcode=1978A\%26A....70..677C&db_key=AST}{{Cavaliere},
  A. \& {Fusco-Femiano}, R.} 1978, \aap, 70, 677

\bibitem[{{Cavaliere} {et~al.}(2009){Cavaliere}, {Lapi}, \&
  {Fusco-Femiano}}]{cavaliere09a}
\href{http://adsabs.harvard.edu/abs/2009ApJ...698..580C}{{Cavaliere}, A.,
  {Lapi}, A., \& {Fusco-Femiano}, R.} 2009, \apj, 698, 580

\bibitem[{{Churazov} {et~al.}(2008){Churazov}, {Forman}, {Vikhlinin},
  {Tremaine}, {Gerhard}, \& {Jones}}]{churazov08a}
\href{http://adsabs.harvard.edu/abs/2008MNRAS.388.1062C}{{Churazov}, E.,
  {Forman}, W., {Vikhlinin}, A., {Tremaine}, S., {Gerhard}, O., \& {Jones}, C.}
  2008, \mnras, 388, 1062

\bibitem[{{Das} {et~al.}(2010){Das}, {Gerhard}, {Churazov}, \&
  {Zhuravleva}}]{das10a}
\href{http://adsabs.harvard.edu/abs/2010MNRAS.409.1362D}{{Das}, P., {Gerhard},
  O., {Churazov}, E., \& {Zhuravleva}, I.} 2010, \mnras, 409, 1362

\bibitem[{{David} {et~al.}(2011){David}, {O'Sullivan}, {Jones}, {Giacintucci},
  {Vrtilek}, {Raychaudhury}, {Nulsen}, {Forman}, {Sun}, \&
  {Donahue}}]{david11a}
\href{http://adsabs.harvard.edu/abs/2011ApJ...728..162D}{{David}, L.~P.,
  {O'Sullivan}, E., {Jones}, C., {Giacintucci}, S., {Vrtilek}, J.,
  {Raychaudhury}, S., {Nulsen}, P.~E.~J., {Forman}, W., {Sun}, M., \&
  {Donahue}, M.} 2011, \apj, 728, 162

\bibitem[{{De Grandi} \& {Molendi}(2001)}]{degrandi01a}
\href{http://adsabs.harvard.edu/abs/2001ApJ...551..153D}{{De Grandi}, S. \&
  {Molendi}, S.} 2001, \apj, 551, 153

\bibitem[{{Donahue} {et~al.}(2006){Donahue}, {Horner}, {Cavagnolo}, \&
  {Voit}}]{donahue06a}
\href{http://adsabs.harvard.edu/cgi-bin/nph-bib_query?bibcode=2006ApJ...643..730D&db_key=AST}{{Donahue},
  M., {Horner}, D.~J., {Cavagnolo}, K.~W., \& {Voit}, G.~M.} 2006, \apj, 643,
  730

\bibitem[{{Dunkley} {et~al.}(2009){Dunkley}, {Komatsu}, {Nolta}, {Spergel},
  {Larson}, {Hinshaw}, {Page}, {Bennett}, {Gold}, {Jarosik}, {Weiland},
  {Halpern}, {Hill}, {Kogut}, {Limon}, {Meyer}, {Tucker}, {Wollack}, \&
  {Wright}}]{dunkley09a}
\href{http://adsabs.harvard.edu/abs/2009ApJS..180..306D}{{Dunkley}, J.,
  {et~al.}} 2009, \apjs, 180, 306

\bibitem[{{Eckert} {et~al.}(2012){Eckert}, {Vazza}, {Ettori}, {Molendi},
  {Nagai}, {Lau}, {Roncarelli}, {Rossetti}, {Snowden}, \&
  {Gastaldello}}]{eckert12a}
\href{http://adsabs.harvard.edu/abs/2012A\%26A...541A..57E}{{Eckert}, D.,
  {Vazza}, F., {Ettori}, S., {Molendi}, S., {Nagai}, D., {Lau}, E.~T.,
  {Roncarelli}, M., {Rossetti}, M., {Snowden}, S.~L., \& {Gastaldello}, F.}
  2012, \aap, 541, A57

\bibitem[{{Evrard} {et~al.}(1996){Evrard}, {Metzler}, \& {Navarro}}]{evrard96a}
\href{http://adsabs.harvard.edu/cgi-bin/nph-bib_query?bibcode=1996ApJ...469..494E&db_key=AST}{{Evrard},
  A.~E., {Metzler}, C.~A., \& {Navarro}, J.~F.} 1996, \apj, 469, 494

\bibitem[{{Eyles} {et~al.}(1991){Eyles}, {Watt}, {Bertram}, {Church}, {Ponman},
  {Skinner}, \& {Willmore}}]{eyles91a}
\href{http://adsabs.harvard.edu/abs/1991ApJ...376...23E}{{Eyles}, C.~J.,
  {Watt}, M.~P., {Bertram}, D., {Church}, M.~J., {Ponman}, T.~J., {Skinner},
  G.~K., \& {Willmore}, A.~P.} 1991, \apj, 376, 23

\bibitem[{{Fabian} {et~al.}(1986){Fabian}, {Thomas}, {White}, \&
  {Fall}}]{fabian86a}
\href{http://adsabs.harvard.edu/cgi-bin/nph-bib_query?bibcode=1986MNRAS.221.1049F&db_key=AST}{{Fabian},
  A.~C., {Thomas}, P.~A., {White}, R.~E., \& {Fall}, S.~M.} 1986, \mnras, 221,
  1049

\bibitem[{{Fang} {et~al.}(2009){Fang}, {Humphrey}, \& {Buote}}]{fang09a}
\href{http://adsabs.harvard.edu/abs/2009ApJ...691.1648F}{{Fang}, T.,
  {Humphrey}, P., \& {Buote}, D.} 2009, \apj, 691, 1648

\bibitem[{{Feroz} {et~al.}(2009){Feroz}, {Hobson}, \& {Bridges}}]{feroz09a}
\href{http://adsabs.harvard.edu/abs/2009MNRAS.398.1601F}{{Feroz}, F., {Hobson},
  M.~P., \& {Bridges}, M.} 2009, \mnras, 398, 1601

\bibitem[{{Finoguenov} {et~al.}(2007){Finoguenov}, {Ponman}, {Osmond}, \&
  {Zimer}}]{finoguenov07a}
\href{http://adsabs.harvard.edu/abs/2007MNRAS.374..737F}{{Finoguenov}, A.,
  {Ponman}, T.~J., {Osmond}, J.~P.~F., \& {Zimer}, M.} 2007, \mnras, 374, 737

\bibitem[{{Frederiksen} {et~al.}(2009){Frederiksen}, {Hansen}, {Host}, \&
  {Roncadelli}}]{fredericksen09a}
\href{http://adsabs.harvard.edu/abs/2009ApJ...700.1603F}{{Frederiksen}, T.~F.,
  {Hansen}, S.~H., {Host}, O., \& {Roncadelli}, M.} 2009, \apj, 700, 1603

\bibitem[{{Gastaldello} {et~al.}(2007{\natexlab{a}}){Gastaldello}, {Buote},
  {Humphrey}, {Zappacosta}, {Brighenti}, \& {Mathews}}]{gastaldello07b}
\href{http://adsabs.harvard.edu/abs/2007hvcg.conf..275G}{{Gastaldello}, F.,
  {Buote}, D.~A., {Humphrey}, P.~J., {Zappacosta}, L., {Brighenti}, F., \&
  {Mathews}, W.~G.} 2007{\natexlab{a}}, in Heating versus Cooling in Galaxies
  and Clusters of Galaxies, ed. H.~{B{\"o}hringer}, G.~W. {Pratt},
  A.~{Finoguenov}, \& P.~{Schuecker}, 275

\bibitem[{{Gastaldello} {et~al.}(2007{\natexlab{b}}){Gastaldello}, {Buote},
  {Humphrey}, {Zappacosta}, {Bullock}, {Brighenti}, \&
  {Mathews}}]{gastaldello07a}
\href{http://adsabs.harvard.edu/abs/2007ApJ...669..158G}{{Gastaldello}, F.,
  {Buote}, D.~A., {Humphrey}, P.~J., {Zappacosta}, L., {Bullock}, J.~S.,
  {Brighenti}, F., \& {Mathews}, W.~G.} 2007{\natexlab{b}}, \apj, 669, 158

\bibitem[{{Gebhardt} \& {Thomas}(2009)}]{gebhardt09a}
\href{http://adsabs.harvard.edu/abs/2009ApJ...700.1690G}{{Gebhardt}, K. \&
  {Thomas}, J.} 2009, \apj, 700, 1690

\bibitem[{{Gnedin} {et~al.}(2004){Gnedin}, {Kravtsov}, {Klypin}, \&
  {Nagai}}]{gnedin04a}
\href{http://adsabs.harvard.edu/cgi-bin/nph-bib_query?bibcode=2004ApJ...616...16G&db_key=AST}{{Gnedin},
  O.~Y., {Kravtsov}, A.~V., {Klypin}, A.~A., \& {Nagai}, D.} 2004, \apj, 616,
  16

\bibitem[{{Gould}(2003)}]{gould03a}
\href{http://adsabs.harvard.edu/abs/2003astro.ph.10577G}{{Gould}, A.} 2003,
  preprint, astro-ph/0310577

\bibitem[{{Hopkins} {et~al.}(2006){Hopkins}, {Hernquist}, {Cox}, {Di Matteo},
  {Robertson}, \& {Springel}}]{hopkins06a}
\href{http://adsabs.harvard.edu/abs/2006ApJS..163....1H}{{Hopkins}, P.~F.,
  {Hernquist}, L., {Cox}, T.~J., {Di Matteo}, T., {Robertson}, B., \&
  {Springel}, V.} 2006, \apjs, 163, 1

\bibitem[{{Humphrey} \& {Buote}(2006)}]{humphrey05a}
\href{http://adsabs.harvard.edu/cgi-bin/nph-bib_query?bibcode=2006ApJ...639..136H&db_key=AST}{{Humphrey},
  P.~J. \& {Buote}, D.~A.} 2006, \apj, 639, 136

\bibitem[{{Humphrey} \& {Buote}(2010)}]{humphrey10a}
\href{http://adsabs.harvard.edu/abs/2010MNRAS.403.2143H}{{Humphrey}, P.~J. \&
  {Buote}, D.~A.} 2010, \mnras, 403, 2143

\bibitem[{{Humphrey} {et~al.}(2012{\natexlab{a}}){Humphrey}, {Buote},
  {Brighenti}, {Flohic}, {Gastaldello}, \& {Mathews}}]{humphrey12a}
\href{http://adsabs.harvard.edu/abs/2012ApJ...748...11H}{{Humphrey}, P.~J.,
  {Buote}, D.~A., {Brighenti}, F., {Flohic}, H.~M.~L.~G., {Gastaldello}, F., \&
  {Mathews}, W.~G.} 2012{\natexlab{a}}, \apj, 748, 11

\bibitem[{{Humphrey} {et~al.}(2008){Humphrey}, {Buote}, {Brighenti},
  {Gebhardt}, \& {Mathews}}]{humphrey08a}
\href{http://adsabs.harvard.edu/abs/2008ApJ...683..161H}{{Humphrey}, P.~J.,
  {Buote}, D.~A., {Brighenti}, F., {Gebhardt}, K., \& {Mathews}, W.~G.} 2008,
  \apj, 683, 161

\bibitem[{{Humphrey} {et~al.}(2009{\natexlab{a}}){Humphrey}, {Buote},
  {Brighenti}, {Gebhardt}, \& {Mathews}}]{humphrey09d}
\href{http://adsabs.harvard.edu/abs/2009ApJ...703.1257H}{{Humphrey}, P.~J.,
  {Buote}, D.~A., {Brighenti}, F., {Gebhardt}, K., \& {Mathews}, W.~G.}
  2009{\natexlab{a}}, \apj, 703, 1257, (H09)

\bibitem[{{Humphrey} {et~al.}(2013){Humphrey}, {Buote}, {Brighenti},
  {Gebhardt}, \& {Mathews}}]{humphrey12c}
\href{http://adsabs.harvard.edu/abs/2013MNRAS.430.1516H}{{Humphrey}, P.~J.,
  {Buote}, D.~A., {Brighenti}, F., {Gebhardt}, K., \& {Mathews}, W.~G.} 2013,
  \mnras, 430, 1516

\bibitem[{{Humphrey} {et~al.}(2011){Humphrey}, {Buote}, {Canizares}, {Fabian},
  \& {Miller}}]{humphrey11a}
\href{http://adsabs.harvard.edu/abs/2011ApJ...729...53H}{{Humphrey}, P.~J.,
  {Buote}, D.~A., {Canizares}, C.~R., {Fabian}, A.~C., \& {Miller}, J.~M.}
  2011, \apj, 729, 53

\bibitem[{{Humphrey} {et~al.}(2006){Humphrey}, {Buote}, {Gastaldello},
  {Zappacosta}, {Bullock}, {Brighenti}, \& {Mathews}}]{humphrey06a}
\href{http://adsabs.harvard.edu/cgi-bin/nph-bib_query?bibcode=2006ApJ...646..899H&db_key=AST}{{Humphrey},
  P.~J., {Buote}, D.~A., {Gastaldello}, F., {Zappacosta}, L., {Bullock}, J.~S.,
  {Brighenti}, F., \& {Mathews}, W.~G.} 2006, \apj, 646, 899, (H06)

\bibitem[{{Humphrey} {et~al.}(2012{\natexlab{b}}){Humphrey}, {Buote},
  {O'Sullivan}, \& {Ponman}}]{humphrey12b}
\href{http://adsabs.harvard.edu/abs/2012ApJ...755..166H}{{Humphrey}, P.~J.,
  {Buote}, D.~A., {O'Sullivan}, E., \& {Ponman}, T.~J.} 2012{\natexlab{b}},
  \apj, 755, 166

\bibitem[{{Humphrey} {et~al.}(2009{\natexlab{b}}){Humphrey}, {Liu}, \&
  {Buote}}]{humphrey09b}
\href{http://adsabs.harvard.edu/abs/2009ApJ...693..822H}{{Humphrey}, P.~J.,
  {Liu}, W., \& {Buote}, D.~A.} 2009{\natexlab{b}}, \apj, 693, 822

\bibitem[{{Jetha} {et~al.}(2007){Jetha}, {Ponman}, {Hardcastle}, \&
  {Croston}}]{jetha07a}
\href{http://adsabs.harvard.edu/abs/2007MNRAS.376..193J}{{Jetha}, N.~N.,
  {Ponman}, T.~J., {Hardcastle}, M.~J., \& {Croston}, J.~H.} 2007, \mnras, 376,
  193

\bibitem[{{Johnson} {et~al.}(2011){Johnson}, {Finoguenov}, {Ponman},
  {Rasmussen}, \& {Sanderson}}]{johnson11a}
\href{http://adsabs.harvard.edu/abs/2011MNRAS.413.2467J}{{Johnson}, R.,
  {Finoguenov}, A., {Ponman}, T.~J., {Rasmussen}, J., \& {Sanderson}, A.~J.~R.}
  2011, \mnras, 413, 2467

\bibitem[{{Johnson} {et~al.}(2009){Johnson}, {Ponman}, \&
  {Finoguenov}}]{johnson09b}
\href{http://adsabs.harvard.edu/abs/2009MNRAS.395.1287J}{{Johnson}, R.,
  {Ponman}, T.~J., \& {Finoguenov}, A.} 2009, \mnras, 395, 1287

\bibitem[{{King}(1972)}]{king72a}
\href{http://adsabs.harvard.edu/abs/1972ApJ...174L.123K}{{King}, I.~R.} 1972,
  \apjl, 174, L123

\bibitem[{{Kravtsov} {et~al.}(2006){Kravtsov}, {Vikhlinin}, \&
  {Nagai}}]{kravtsov06a}
\href{http://adsabs.harvard.edu/abs/2006ApJ...650..128K}{{Kravtsov}, A.~V.,
  {Vikhlinin}, A., \& {Nagai}, D.} 2006, \apj, 650, 128

\bibitem[{{Lewis} {et~al.}(2002){Lewis}, {Stocke}, \& {Buote}}]{lewis02a}
\href{http://adsabs.harvard.edu/cgi-bin/nph-bib_query?bibcode=2002ApJ...573L..13L&db_key=AST}{{Lewis},
  A.~D., {Stocke}, J.~T., \& {Buote}, D.~A.} 2002, \apjl, 573, L13

\bibitem[{{Liu} {et~al.}(2012{\natexlab{a}})}]{liu12a}
{Liu}, W. {et~al.} 2012{\natexlab{a}}, in preparation

\bibitem[{{Liu} {et~al.}(2012{\natexlab{b}})}]{liu12c}
---. 2012{\natexlab{b}}, in preparation

\bibitem[{{Lloyd-Davies} {et~al.}(2000){Lloyd-Davies}, {Ponman}, \&
  {Cannon}}]{lloyddavies00a}
\href{http://adsabs.harvard.edu/cgi-bin/nph-bib_query?bibcode=2000MNRAS.315..689L&db_key=AST}{{Lloyd-Davies},
  E.~J., {Ponman}, T.~J., \& {Cannon}, D.~B.} 2000, \mnras, 315, 689

\bibitem[{{Mahdavi} {et~al.}(2005){Mahdavi}, {Finoguenov}, {B{\"o}hringer},
  {Geller}, \& {Henry}}]{mahdavi05a}
\href{http://adsabs.harvard.edu/abs/2005ApJ...622..187M}{{Mahdavi}, A.,
  {Finoguenov}, A., {B{\"o}hringer}, H., {Geller}, M.~J., \& {Henry}, J.~P.}
  2005, \apj, 622, 187

\bibitem[{{Mahdavi} {et~al.}(2008){Mahdavi}, {Hoekstra}, {Babul}, \&
  {Henry}}]{mahdavi08a}
\href{http://adsabs.harvard.edu/abs/2008MNRAS.384.1567M}{{Mahdavi}, A.,
  {Hoekstra}, H., {Babul}, A., \& {Henry}, J.~P.} 2008, \mnras, 384, 1567

\bibitem[{{Mahdavi} {et~al.}(2007){Mahdavi}, {Hoekstra}, {Babul}, {Sievers},
  {Myers}, \& {Henry}}]{mahdavi07a}
\href{http://adsabs.harvard.edu/abs/2007ApJ...664..162M}{{Mahdavi}, A.,
  {Hoekstra}, H., {Babul}, A., {Sievers}, J., {Myers}, S.~T., \& {Henry},
  J.~P.} 2007, \apj, 664, 162

\bibitem[{{Mathews}(1978)}]{mathews78a}
\href{http://adsabs.harvard.edu/abs/1978ApJ...219..413M}{{Mathews}, W.~G.}
  1978, \apj, 219, 413

\bibitem[{{Mathews} \& {Brighenti}(2003)}]{mathews03a}
\href{http://adsabs.harvard.edu/cgi-bin/nph-bib_query?bibcode=2003ARA\%26A..41..191M&amp;db_key=AST}{{Mathews},
  W.~G. \& {Brighenti}, F.} 2003, \araa, 41, 191

\bibitem[{{Mathews} \& {Guo}(2011)}]{mathews11a}
\href{http://adsabs.harvard.edu/abs/2011ApJ...738..155M}{{Mathews}, W.~G. \&
  {Guo}, F.} 2011, \apj, 738, 155

\bibitem[{{Mazzotta} {et~al.}(2004){Mazzotta}, {Rasia}, {Moscardini}, \&
  {Tormen}}]{mazzotta04a}
\href{http://adsabs.harvard.edu/abs/2004MNRAS.354...10M}{{Mazzotta}, P.,
  {Rasia}, E., {Moscardini}, L., \& {Tormen}, G.} 2004, \mnras, 354, 10

\bibitem[{{McCarthy} {et~al.}(2010){McCarthy}, {Schaye}, {Ponman}, {Bower},
  {Booth}, {Dalla Vecchia}, {Crain}, {Springel}, {Theuns}, \&
  {Wiersma}}]{mccarthy10a}
\href{http://adsabs.harvard.edu/abs/2010MNRAS.406..822M}{{McCarthy}, I.~G.,
  {Schaye}, J., {Ponman}, T.~J., {Bower}, R.~G., {Booth}, C.~M., {Dalla
  Vecchia}, C., {Crain}, R.~A., {Springel}, V., {Theuns}, T., \& {Wiersma},
  R.~P.~C.} 2010, \mnras, 406, 822

\bibitem[{{Miller} {et~al.}(2012){Miller}, {Bautz}, {George}, {Mushotzky},
  {Davis}, \& {Henry}}]{miller11a}
\href{http://adsabs.harvard.edu/abs/2012AIPC.1427...13M}{{Miller}, E.~D.,
  {Bautz}, M., {George}, J., {Mushotzky}, R., {Davis}, D., \& {Henry}, J.~P.}
  2012, in American Institute of Physics Conference Series, Vol. 1427, American
  Institute of Physics Conference Series, ed. R.~{Petre}, K.~{Mitsuda}, \&
  L.~{Angelini}, 13--20

\bibitem[{{Molendi}(2002)}]{molendi02a}
\href{http://adsabs.harvard.edu/abs/2002ApJ...580..815M}{{Molendi}, S.} 2002,
  \apj, 580, 815

\bibitem[{{Molendi} \& {Pizzolato}(2001)}]{molendi01b}
\href{http://adsabs.harvard.edu/cgi-bin/nph-bib_query?bibcode=2001ApJ...560..194M&db_key=AST}{{Molendi},
  S. \& {Pizzolato}, F.} 2001, \apj, 560, 194

\bibitem[{{Nagai} {et~al.}(2007){Nagai}, {Vikhlinin}, \& {Kravtsov}}]{nagai07a}
\href{http://adsabs.harvard.edu/abs/2007ApJ...655...98N}{{Nagai}, D.,
  {Vikhlinin}, A., \& {Kravtsov}, A.~V.} 2007, \apj, 655, 98

\bibitem[{{Navarro} {et~al.}(1997){Navarro}, {Frenk}, \& {White}}]{navarro97}
\href{http://adsabs.harvard.edu/cgi-bin/nph-bib_query?bibcode=1997ApJ...490..493N&db_key=AST}{{Navarro},
  J.~F., {Frenk}, C.~S., \& {White}, S.~D.~M.} 1997, \apj, 490, 493

\bibitem[{{Okabe} {et~al.}(2010){Okabe}, {Takada}, {Umetsu}, {Futamase}, \&
  {Smith}}]{okabe10a}
\href{http://adsabs.harvard.edu/abs/2010PASJ...62..811O}{{Okabe}, N., {Takada},
  M., {Umetsu}, K., {Futamase}, T., \& {Smith}, G.~P.} 2010, \pasj, 62, 811

\bibitem[{Pearson(1900)}]{pearson1900a}
\href{http://www.tandfonline.com/doi/abs/10.1080/14786440009463897}{Pearson,
  K.} 1900, Philosophical Magazine Series 5, 50, 157

\bibitem[{{Peterson} {et~al.}(2003){Peterson}, {Kahn}, {Paerels}, {Kaastra},
  {Tamura}, {Bleeker}, {Ferrigno}, \& {Jernigan}}]{peterson03a}
\href{http://adsabs.harvard.edu/cgi-bin/nph-bib_query?bibcode=2003ApJ...590..207P&db_key=AST}{{Peterson},
  J.~R., {Kahn}, S.~M., {Paerels}, F.~B.~S., {Kaastra}, J.~S., {Tamura}, T.,
  {Bleeker}, J.~A.~M., {Ferrigno}, C., \& {Jernigan}, J.~G.} 2003, \apj, 590,
  207

\bibitem[{{Piffaretti} \& {Valdarnini}(2008)}]{piffaretti08a}
\href{http://adsabs.harvard.edu/abs/2008A\%26A...491...71P}{{Piffaretti}, R. \&
  {Valdarnini}, R.} 2008, \aap, 491, 71

\bibitem[{{Pizzolato} {et~al.}(2003){Pizzolato}, {Molendi}, {Ghizzardi}, \& {De
  Grandi}}]{pizzolato03a}
\href{http://adsabs.harvard.edu/abs/2003ApJ...592...62P}{{Pizzolato}, F.,
  {Molendi}, S., {Ghizzardi}, S., \& {De Grandi}, S.} 2003, \apj, 592, 62

\bibitem[{{Planck Collaboration} {et~al.}(2011){Planck Collaboration},
  {Aghanim}, {Arnaud}, {Ashdown}, {Aumont}, {Baccigalupi}, {Balbi}, {Banday},
  {Barreiro}, {Bartelmann}, {Bartlett}, {Battaner}, {Benabed}, {Beno{\^i}t},
  {Bernard}, {Bersanelli}, {Bhatia}, {Bock}, {Bonaldi}, {Bond}, {Borrill},
  {Bouchet}, {Brown}, {Bucher}, {Burigana}, {Cabella}, {Cardoso}, {Catalano},
  {Cay{\'o}n}, {Challinor}, {Chamballu}, {Chary}, {Chiang}, {Chiang}, {Chon},
  {Christensen}, {Churazov}, {Clements}, {Colafrancesco}, {Colombi}, {Couchot},
  {Coulais}, {Crill}, {Cuttaia}, {da Silva}, {Dahle}, {Danese}, {de Bernardis},
  {de Gasperis}, {de Rosa}, {de Zotti}, {Delabrouille}, {Delouis},
  {D{\'e}sert}, {Diego}, {Dolag}, {Donzelli}, {Dor{\'e}}, {D{\"o}rl},
  {Douspis}, {Dupac}, {Efstathiou}, {En{\ss}lin}, {Finelli}, {Flores-Cacho},
  {Forni}, {Frailis}, {Franceschi}, {Fromenteau}, {Galeotta}, {Ganga},
  {G{\'e}nova-Santos}, {Giard}, {Giardino}, {Giraud-H{\'e}raud},
  {Gonz{\'a}lez-Nuevo}, {G{\'o}rski}, {Gratton}, {Gregorio}, {Gruppuso},
  {Harrison}, {Henrot-Versill{\'e}}, {Hern{\'a}ndez-Monteagudo}, {Herranz},
  {Hildebrandt}, {Hivon}, {Hobson}, {Holmes}, {Hovest}, {Hoyland},
  {Huffenberger}, {Jaffe}, {Jones}, {Juvela}, {Keih{\"a}nen}, {Keskitalo},
  {Kisner}, {Kneissl}, {Knox}, {Kurki-Suonio}, {Lagache}, {Lamarre}, {Lasenby},
  {Laureijs}, {Lawrence}, {Leach}, {Leonardi}, {Linden-V{\o}rnle},
  {L{\'o}pez-Caniego}, {Lubin}, {Mac{\'{\i}}as-P{\'e}rez}, {MacTavish},
  {Maffei}, {Maino}, {Mandolesi}, {Mann}, {Maris}, {Marleau},
  {Mart{\'{\i}}nez-Gonz{\'a}lez}, {Masi}, {Matarrese}, {Matthai}, {Mazzotta},
  {Melchiorri}, {Melin}, {Mendes}, {Mennella}, {Mitra},
  {Miville-Desch{\^e}nes}, {Moneti}, {Montier}, {Morgante}, {Mortlock},
  {Munshi}, {Murphy}, {Naselsky}, {Natoli}, {Netterfield},
  {N{\o}rgaard-Nielsen}, {Noviello}, {Novikov}, {Novikov}, {Osborne}, {Pajot},
  {Pasian}, {Patanchon}, {Perdereau}, {Perotto}, {Perrotta}, {Piacentini},
  {Piat}, {Pierpaoli}, {Piffaretti}, {Plaszczynski}, {Pointecouteau},
  {Polenta}, {Ponthieu}, {Poutanen}, {Pratt}, {Pr{\'e}zeau}, {Prunet}, {Puget},
  {Rebolo}, {Reinecke}, {Renault}, {Ricciardi}, {Riller}, {Ristorcelli},
  {Rocha}, {Rosset}, {Rubi{\~n}o-Mart{\'{\i}}n}, {Rusholme}, {Sandri},
  {Santos}, {Schaefer}, {Scott}, {Seiffert}, {Smoot}, {Starck}, {Stivoli},
  {Stolyarov}, {Sunyaev}, {Sygnet}, {Tauber}, {Terenzi}, {Toffolatti},
  {Tomasi}, {Tristram}, {Tuovinen}, {Valenziano}, {Vibert}, {Vielva}, {Villa},
  {Vittorio}, {Wandelt}, {White}, {White}, {Yvon}, {Zacchei}, \&
  {Zonca}}]{planck11a}
\href{http://adsabs.harvard.edu/abs/2011A\%26A...536A..10P}{{Planck
  Collaboration}, {et~al.}} 2011, \aap, 536, A10

\bibitem[{{Ponman} {et~al.}(1999){Ponman}, {Cannon}, \& {Navarro}}]{ponman99a}
\href{http://adsabs.harvard.edu/cgi-bin/nph-bib_query?bibcode=1999Natur.397..135P&db_key=AST}{{Ponman},
  T.~J., {Cannon}, D.~B., \& {Navarro}, J.~F.} 1999, \nat, 397, 135

\bibitem[{{Pratt} {et~al.}(2010){Pratt}, {Arnaud}, {Piffaretti},
  {B{\"o}hringer}, {Ponman}, {Croston}, {Voit}, {Borgani}, \&
  {Bower}}]{pratt10a}
\href{http://adsabs.harvard.edu/abs/2010A\%26A...511A..85P}{{Pratt}, G.~W.,
  {Arnaud}, M., {Piffaretti}, R., {B{\"o}hringer}, H., {Ponman}, T.~J.,
  {Croston}, J.~H., {Voit}, G.~M., {Borgani}, S., \& {Bower}, R.~G.} 2010,
  \aap, 511, A85+

\bibitem[{{Sanderson} {et~al.}(2003){Sanderson}, {Ponman}, {Finoguenov},
  {Lloyd-Davies}, \& {Markevitch}}]{sanderson03a}
\href{http://adsabs.harvard.edu/abs/2003MNRAS.340..989S}{{Sanderson}, A.~J.~R.,
  {Ponman}, T.~J., {Finoguenov}, A., {Lloyd-Davies}, E.~J., \& {Markevitch},
  M.} 2003, \mnras, 340, 989

\bibitem[{{Silk} \& {Rees}(1998)}]{silk98a}
\href{http://adsabs.harvard.edu/abs/1998A\%26A...331L...1S}{{Silk}, J. \&
  {Rees}, M.~J.} 1998, \aap, 331, L1

\bibitem[{{Simionescu} {et~al.}(2011){Simionescu}, {Allen}, {Mantz}, {Werner},
  {Takei}, {Morris}, {Fabian}, {Sanders}, {Nulsen}, {George}, \&
  {Taylor}}]{simionescu11a}
\href{http://adsabs.harvard.edu/abs/2011arXiv1102.2429S}{{Simionescu}, A.,
  {et~al.}} 2011, Science, in press (arXiv:1102.2429)

\bibitem[{{Sun}(2012)}]{sun12a}
\href{http://adsabs.harvard.edu/abs/2012NJPh...14d5004S}{{Sun}, M.} 2012, New
  Journal of Physics, 14, 045004

\bibitem[{{Sun} {et~al.}(2009){Sun}, {Voit}, {Donahue}, {Jones}, {Forman}, \&
  {Vikhlinin}}]{sun09a}
\href{http://adsabs.harvard.edu/abs/2009ApJ...693.1142S}{{Sun}, M., {Voit},
  G.~M., {Donahue}, M., {Jones}, C., {Forman}, W., \& {Vikhlinin}, A.} 2009,
  \apj, 693, 1142

\bibitem[{{Tozzi} \& {Norman}(2001)}]{tozzi01a}
\href{http://adsabs.harvard.edu/cgi-bin/nph-bib_query?bibcode=2001ApJ...546...63T&db_key=AST}{{Tozzi},
  P. \& {Norman}, C.} 2001, \apj, 546, 63

\bibitem[{{Urban} {et~al.}(2011){Urban}, {Werner}, {Simionescu}, {Allen}, \&
  {B{\"o}hringer}}]{urban11a}
\href{http://adsabs.harvard.edu/abs/2011MNRAS.414.2101U}{{Urban}, O., {Werner},
  N., {Simionescu}, A., {Allen}, S.~W., \& {B{\"o}hringer}, H.} 2011, \mnras,
  414, 2101

\bibitem[{{Vikhlinin}(2006)}]{vikhlinin06c}
\href{http://adsabs.harvard.edu/abs/2006ApJ...640..710V}{{Vikhlinin}, A.} 2006,
  \apj, 640, 710

\bibitem[{{Vikhlinin} {et~al.}(2005){Vikhlinin}, {Markevitch}, {Murray},
  {Jones}, {Forman}, \& {Van Speybroeck}}]{vikhlinin05a}
\href{http://adsabs.harvard.edu/cgi-bin/nph-bib_query?bibcode=2005ApJ...628..655V&db_key=AST}{{Vikhlinin},
  A., {Markevitch}, M., {Murray}, S.~S., {Jones}, C., {Forman}, W., \& {Van
  Speybroeck}, L.} 2005, \apj, 628, 655

\bibitem[{{Vikhlinin} {et~al.}(2009){Vikhlinin}, {Kravtsov}, {Burenin},
  {Ebeling}, {Forman}, {Hornstrup}, {Jones}, {Murray}, {Nagai}, {Quintana}, \&
  {Voevodkin}}]{vikhlinin09a}
\href{http://adsabs.harvard.edu/abs/2009ApJ...692.1060V}{{Vikhlinin}, A.,
  {et~al.}} 2009, \apj, 692, 1060

\bibitem[{{Voit}(2005)}]{voit05c}
\href{http://adsabs.harvard.edu/abs/2005RvMP...77..207V}{{Voit}, G.~M.} 2005,
  Reviews of Modern Physics, 77, 207

\bibitem[{{Voit} \& {Donahue}(2005)}]{voit05a}
\href{http://adsabs.harvard.edu/cgi-bin/nph-bib_query?bibcode=2005ApJ...634..955V&db_key=AST}{{Voit},
  G.~M. \& {Donahue}, M.} 2005, \apj, 634, 955

\bibitem[{{Voit} {et~al.}(2005){Voit}, {Kay}, \& {Bryan}}]{voit05b}
\href{http://adsabs.harvard.edu/abs/2005MNRAS.364..909V}{{Voit}, G.~M., {Kay},
  S.~T., \& {Bryan}, G.~L.} 2005, \mnras, 364, 909

\bibitem[{{Werner} {et~al.}(2012){Werner}, {Allen}, \&
  {Simionescu}}]{werner12a}
\href{http://adsabs.harvard.edu/abs/2012arXiv1205.1563W}{{Werner}, N., {Allen},
  S.~W., \& {Simionescu}, A.} 2012, \mnras, in press (arXiv:1205.1563)

\bibitem[{{Wong} {et~al.}(2011){Wong}, {Irwin}, {Yukita}, {Million}, {Mathews},
  \& {Bregman}}]{wong11a}
\href{http://adsabs.harvard.edu/abs/2011ApJ...736L..23W}{{Wong}, K.-W.,
  {Irwin}, J.~A., {Yukita}, M., {Million}, E.~T., {Mathews}, W.~G., \&
  {Bregman}, J.~N.} 2011, \apjl, 736, L23

\end{thebibliography}


\end{document}